\documentclass[twocolumn,a4paper,superscriptaddress,floatfix,longbibliography]{revtex4}
\usepackage{graphicx,amsfonts,amsmath,bm,color,tabularx}
\usepackage{multirow}
\usepackage[normalem]{ulem}

\newcommand{\I}{{\rm i}}
\newcommand{\no}{\mathfrak{n}}
\newcommand{\mob}{\bm{\mathfrak{m}}}
\newcommand{\mo}{\mathfrak{m}}

\begin{document}

\title{Spin-orbit coupling induced orbital entanglement in a three-band Hubbard model}

\author{Petr A. Igoshev}
\affiliation{Institute of Metal Physics, S. Kovalevskaya str. 18, 620108 Ekaterinburg, Russia}

\author{Danil E. Chizhov}
\affiliation{Department of Theoretical Physics and Applied Mathematics, Ural Federal University, Mira str. 19, 620002 Ekaterinburg, Russia}

\author{Valentin Yu. Irkhin}
\affiliation{Institute of Metal Physics, S. Kovalevskaya str. 18, 620108 Ekaterinburg, Russia}

\author{Sergey V. Streltsov}
\affiliation{Institute of Metal Physics, S. Kovalevskaya str. 18, 620108 Ekaterinburg, Russia}
\affiliation{Department of Theoretical Physics and Applied Mathematics, Ural Federal University, Mira str. 19, 620002 Ekaterinburg, Russia}

\date{\today}

\date{\today}
\begin{abstract}
The effect of the spin-orbit coupling on the ground state properties of the square-lattice three-band Hubbard model with a single electron per site is studied by a generalized Hartree-Fock approximation. We calculate the full phase diagram and show that there appear additional orbital-entangled phases brought about by competition of various exchange channels or by the spin-orbit coupling in addition to conventional states stabilized by the Kugel-Khomskii mechanism. One of these phases  previously proposed to explain magnetic properties of Sr$_2$VO$_4$  is characterized by vanishing dipolar magnetic moments and antiferro-octupolar ordering. We calculated microscopic parameters for this material and demonstrate that it is located near a phase boundary of two orbital-entangled and two conventional antiferromagnetic ferro-orbital states.

\end{abstract}

%\keywords{}
\maketitle

%%%%%%%%%%%%%%%%%%%%%%%%%%%%%%%%%%%%%%
\section{Introduction}

The Hubbard Hamiltonian on a square lattice has become not only one of the most studied models over the past decades, but has turned out to be a standard test bed for various theoretical methods. Nevertheless, this model still harbors  intriguing physics to be uncovered. Two ingredients --- orbital degeneracy and the spin-orbit coupling --- substantially enrich the variety of physical phenomena described by this model. 

Being a very important theoretical concept, the model is extremely useful for practical applications, and not only in connection with high-temperature superconductivity of cuprates, but for many other materials and phenomena including the Kugel-Khomskii mechanism of orbital ordering in K$_2$CuF$_4$~\cite{Khomskii1973,KK-UFN}, unconventional superconductivity in Sr$_2$RuO$_4$~\cite{nelson2004odd}, orbital-selective physics in Ca$_2$RuO$_4$~\cite{Anisimov2002a} and the spin-orbit assisted Mott transition in Sr$_2$IrO$_4$~\cite{Kim2008}. There are also other layered perovskites with transition metals forming a square lattice, which demonstrate intriguing and yet to be understood physical properties. Examples include the anomalous staircase field dependence of magnetization~\cite{Li2016} together with half metallicity~\cite{Matsuno2004,Pandey2010,Wu2012} and obscure spin state of Co (electronic configuration $3d^5$) in Sr$_2$CoO$_4$~\cite{Wang2005T,Lee2006,Pandey2010,Wu2012,bhardwaj2024}, or Sr$_2$CrO$_4$ ($3d^2$) with a reversed crystal field~\cite{ishikawa2017}, strong interplay between spin and orbital degrees of freedom~\cite{Pandey2021}, and possible formation of orbitally ordered states switchable by ultrafast optical spectroscopy~\cite{lee2022} and destroyable by pressure~\cite{Yamauchi2019}.

Another example is Sr$_2$VO$_4$ with V$^{4+}$ ions having ionic configuration $3d^1$. One might expect formation of a long-range magnetic order at low enough temperature and, indeed, there is an anomaly in magnetic susceptibility at $\sim 100$~K, but neutron measurements do not detect any magnetic moment even at 5~K~\cite{Cyrot1990}. $\mu$SR experiments evidence formation of an antiferromagnetic order below 8~K~\cite{Sugiyama2014}. Various theoretical models have been proposed to resolve the problem of vanishing local magnetic moment in this material. In particular, Imai {\it et al.} found a complicated spin-orbital order and severe competition between various magnetic/orbital configurations~\cite{Imai2005}. Jackeli and Khaliullin put forward an idea of a hidden magnetic order, when orbital and spin moments are reduced to zero at each lattice site and  magnetic octupole order instead  ~\cite{Jackeli2009a}. The Jackeli--Khaliullin state is characterized by vanishing dipolar magnetic moment and antiferro-octupolar order with {\it two} nonzero octupolar moments transformed by two nonequivalent representations. Eremin {\it et al.} suggested an alternative state with nonvanishing but compensating each other orbital and spin moments~\cite{Eremin2011}. Density functional theory calculations by Kim {\it et al} stress the importance of frustration effects and argue that  spin-liquid or spin-glass states can be realized at very low temperatures~\cite{Kim2017b}.

In the present paper we perform a detailed study of a three-orbital Hubbard model on the square lattice with a checkerboard order taking into account the spin-orbit coupling and tetragonal crystal-field splitting using a generalized  Hartree-Fock approximation (HFA). Particular attention is paid to the situation of a single $d$ electron, being characteristic for Sr$_2$VO$_4$. The ground-state phase diagram and physical properties of each phase are discussed in detail. We show that Sr$_2$VO$_4$ is in a region of the phase diagram where two highly unusual states with orbital-entangled wave functions (one of which is characterized by zero dipolar but finite octupolar magnetic moment) and two more conventional states with ferromagnetic (but different antiferro-orbitally ordered configurations) are realized. 
\begin{figure}[t]
   \centering
  \includegraphics[width=0.45\textwidth]{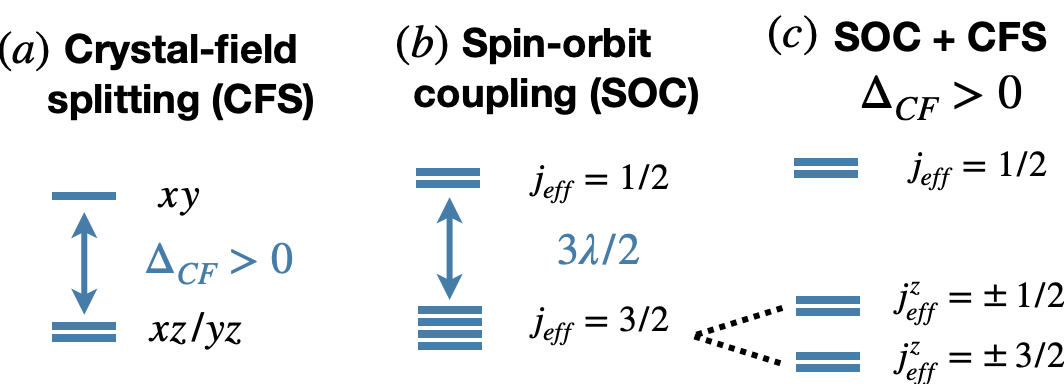}
  \caption{\label{Fig:levels} Illustration of effects of (a) the crystal-field splitting given by \eqref{eq:H_CF} (positive $\Delta_{\rm CF}$ corresponds to elongation of octahedra surrounding transition metals), (b) the spin-orbit coupling \eqref{eq:H_so}, and (c) their combined action in the case of  $\lambda > \Delta_{\rm CF} >0$, which shifts states of $|j_{\rm eff} = 3/2, j_{\rm eff}^z = \pm 1/2 \rangle$ up in energy.}
\end{figure}

%%%%%%%%%%%%%%%%%%%%%%%%%%%%%%%%%%%%%%
\section{Model and method}\label{sec:model_method}
The Hamiltonian of our model reads 
\begin{equation}\label{eq:hamilt}
\mathcal{H} = \mathcal{H}_{\rm t} + \mathcal{H}_{\rm CF} + \mathcal{H}_{\rm so} + \mathcal{H}_{\rm Coulomb},
\end{equation}
where the first term describes intersite hopping processes
\begin{equation}\label{eq:H_tr_def}
	\mathcal{H}_{\rm t} = \sum_{ijmm'\sigma}t^{mm'}_{ij}c^\dag_{im\sigma}c^{}_{jm'\sigma},
\end{equation}
where $m, m' = xz, xy, yz$ numerate orbitals, $\sigma, \sigma' = \uparrow, \downarrow$ define spin projection, and $t_{ij}^{mm'}$ are hopping amplitudes between corresponding orbitals on sites $i$ and $j$. 
The second term
\begin{equation}\label{eq:H_CF}
\mathcal{H}_{\rm CF} = \Delta_{\rm CF}\sum_{i\sigma}c^\dag_{i,xy,\sigma}c^{}_{i,xy,\sigma}
\end{equation}
sets up tetragonal crystal-field splitting $\Delta_{\rm CF}$. The third term in~Eq.~\eqref{eq:hamilt} describes the spin-orbit coupling with the strength defined by the corresponding parameter $\lambda$,
\begin{equation}\label{eq:H_so}
\mathcal{H}_{\rm so} = -\lambda\sum_{i,mm'}\bm{\mathcal{S}}_{i;mm'}\cdot \bm{l}_{mm'}.
\end{equation}
Here three components $o=1,2,3$ of 
\begin{equation}\label{eq:S_mm'_def}
\mathcal{S}^{(o)}_{i;mm'} = \frac12\sum_{\sigma\sigma'}c^\dag_{im\sigma}\sigma^o_{\sigma\sigma'}c^{}_{im'\sigma'}  
\end{equation}
give (generalized) spin operators, and $o=0$ provides information about 
inter-orbital occupation, $\sigma^o$ are Pauli matrices, and $\mathbf{l}_{mm'}$ are conventional  matrices of orbital momentum in the  basis of cubic harmonics
\begin{eqnarray}
\label{eq:l_def_xy}
l^x_{mm'} &=& 
\left(
\begin{matrix}
0 & -\I & 0\\
+\I & 0 & 0\\
0 & 0 & 0
\end{matrix}
\right), 
l^y_{mm'} = 
\left(
\begin{matrix}
0 & 0 & 0\\
0 & 0 & -\I\\
0 & +\I & 0
\end{matrix}
\right), \\
%\end{equation}
%\begin{equation}
\label{eq:l_def_z}
    l^z_{mm'} &=& 
\left(\begin{matrix}
0 & 0 & +\I\\
0 & 0 & 0\\
-\I & 0 & 0
\end{matrix}
\right).
\end{eqnarray}
The minus sign in Eq.~\eqref{eq:H_so} is needed to work with the $t_{2g}$ orbitals, which can be modeled as effective $p$ orbitals ($l = 1$), but with opposite sign of the spin-orbit coupling constant; see, e.g.,~\cite{Abragam,Streltsov2017Chem}.

The on-site Coulomb interaction is taken in the following form:
\begin{multline}\label{eq:main_H_Coulomb}
\mathcal{H}_{\rm Coulomb} = \frac{U}2\sum_{im\sigma}n_{im\sigma}n_{im\bar\sigma} + \frac{U'}2\sum_{i,m\neq m'}n_{im}n_{im'}  \\
- \frac{J_{\rm H}}2\sum_{im\ne m';\sigma\sigma'}c^\dag_{i,m\sigma}c^{}_{i,m\sigma'}c^\dag_{i,m'\sigma'}c^{}_{i,m'\sigma}
\\
-\frac{J_{\rm d}}2\sum_{im\ne m';\sigma}c^\dag_{i,m\sigma}c^{}_{i,m'\bar\sigma}c^\dag_{i,m\bar\sigma}c^{}_{i,m'\sigma},
\end{multline}
where $U(U')$ is the on-site Coulomb intraorbital~(interorbital) interaction parameter, $J_{\rm H}$ is the Hund's intra-atomic exchange (in the Kanamori representation~\cite{Kanamori1963} $U = U' + 2J_{\rm H}$), pair-hopping matrix element $J_{\rm d} = J_{\rm H}$ is a~real number, since we work with real cubic ($t_{\rm 2g}$) orbitals, and
$n_{im\sigma} = c^\dag_{im\sigma}c^{}_{im\sigma}$ and $n_{im} = \sum_\sigma n_{im\sigma}$ are the occupation number operators. The notation $\bar\sigma = -\sigma$ is used.

Below we consider the case of a single electron per lattice site under the assumption that a two-sublattice checkerboard long-range order is established. In the case of a Mott-Hubbard insulator with an integer band filling the Hartree-Fock approximation provides reasonable results \cite{irkhin1989spin}. Indeed, in this case HFA yields correct energy values in the atomic limit, provided that virtual states are treated correctly within this method (i.e., essentially Anderson's kinetic exchange effects).   
This approximation was also successfully applied to describe the electron and magnon spectrum \cite{irkhin1989spin} and magnetic phase diagram of the two-band $s-d$ exchange model~\cite{pankratova2021incommensurate}.

Otherwise, HFA gives only a qualitative estimate of ground state energy missing a vertex correction to virtual state energies. However, the quality of HFA indeed depends on the phases considered and an estimation of its applicability is not straightforward.

Due to the local character of the  Coulomb interaction Hamiltonian \eqref{eq:main_H_Coulomb}, the generalized  HFA is fully specified by a \textit{local order} characterized by a correlator
\begin{equation}\label{eq:C_local_def}
C^i_{m\sigma;m'\sigma'}\equiv\langle c^\dag_{im\sigma}c^{}_{im'\sigma'}\rangle.
\end{equation}
See details of the derivation in Appendix~B; see Eq.~\eqref{eq:H_int^HFA[W]}.

We consider the decomposition of the correlator into a complete set of Pauli matrices
\begin{equation}\label{eq:Pauli_decomposition}
C^{i}_{m\sigma;m'\sigma'} = \no^i_{mm'}\sigma^0_{\sigma'\sigma} +  \mob^{i}_{mm'}\cdot{\bm\sigma}_{\sigma'\sigma},
\end{equation}
where
\begin{eqnarray}
\label{eq:n_mm_def}
\no^i_{mm'} &=& \left\langle \mathcal{S}^{(0)}_{i;mm'}\right\rangle, \\
\label{eq:m_mm_def}
\mob^i_{mm'} &=& \left\langle {\bm{\mathcal{S}}}_{i;mm'}\right\rangle
\end{eqnarray}
are the components of 4-component (charge-spin) vector $N^i_{mm'}(\no^i_{mm'},\mob^i_{mm'})$. 
%and $C^i_{m;m'}$ should be meant as a matrix with respect to indices~$\sigma$ and~$\sigma'$, t means matrix transpose. 

Fourier transform of an arbitrary field $\varphi^i$ can be defined in a standard way  $\varphi(\mathbf{q}) = (1/N)\sum_{i}\exp(\I\mathbf{qR}_i)\varphi^i$,  where $N$ is a number of lattice sites and $\mathbf{q}$ is a~wave vector.  For two-sublattice valued $\varphi^i$ one can once again single out two components of a field $\varphi$
\begin{equation}\label{eq:phi_two_terms}
\varphi(\mathbf{q}) = \delta_{\mathbf{q}0}\varphi^{\rm u} + \delta_{\mathbf{qQ}}\varphi^{\rm s}, 
\end{equation}
where $\varphi^{\rm u}$~($\varphi^{\rm s}$) is the uniform (staggered) component. We restrict ourselves to considering a two-sublattice ordering and take $\varphi^i = C^i_{m\sigma;m'\sigma'}$ with two components, a uniform ($C^{\rm u}$) and staggered ($C^{\rm s}$): 
\begin{equation}\label{eq:AFM_general}
C^i_{m\sigma;m'\sigma'} = C^{\rm u}_{m\sigma;m'\sigma'} + \exp[\I\mathbf{QR}_i]C^{\rm s}_{m\sigma;m'\sigma'}
\end{equation}
(from a mathematical point of view, this limits a class of the general HFA equation solutions). 
This approach provides all two-sublattice long-range order solutions for within HFA. 
Thereby, we take into account that the ordering has only two components (uniform and staggered), see~Eqs.~(\ref{eq:phi_two_terms}) and~(\ref{eq:AFM_general}), so that the Fourier transforms of $\no^i_{mm'}$ and $\mob^i_{mm'}$ have only $\mathbf{q} = \mathbf{0}, \mathbf{Q}$ nonzero contributions~(see~Appendix~B).  

Within the Hartree-Fock approximation, the total  Hamiltonian reads 
\begin{multline}\label{eq:H_MF_total}
\mathcal{H}_{\rm MF} = \sum_{\mathbf{k}_1\mathbf{k}'_1;mm'}\sum_{\sigma\sigma'}\left(
\left[\varepsilon_{mm'}(\mathbf{k}_1)\delta_{\sigma\sigma'} + \mathcal{F}^{(0)\rm u}_{mm'}\delta_{\sigma\sigma'} 
\right.
\right.
\\
\left.
\left.
-{\bm{\mathcal{F}}}^{\rm u}_{mm'}\bm{\sigma}_{\sigma\sigma'} - (\lambda/2)\mathbf{l}_{mm'}\bm{\sigma}_{\sigma\sigma'}\right]\delta_{\mathbf{k}_1\mathbf{k}'_1} 
\right.
\\
+\left. 
\left[{\mathcal{F}^{(0)\rm s}_{mm'}}\delta_{\sigma\sigma'} - {\bm{\mathcal{F}}}^{\rm s}_{mm'}\cdot\bm{\sigma}_{\sigma\sigma'}\right]\delta_{\mathbf{k}_1,\mathbf{k}'_1+\mathbf{Q}}\right)
c^\dag_{\mathbf{k}_1m\sigma}c^{}_{\mathbf{k}'_1m'\sigma'},
\end{multline}
where mean fields $\mathcal{F}^{(0)i}_{mm'}$ and $\bm{\mathcal{F}}^i_{mm'}$ generally have complex orbital structure
\begin{eqnarray}
\label{eq:Fn^i_def}
\mathcal{F}^{(0)i}_{mm'} &=& -(U' - 2J_{\rm H})\no^i_{m'm} + J_{\rm d}\no^i_{mm'}
\\
\nonumber
&& +\delta_{mm'}(2U' - J_{\rm H})\sum_{m''}\no^i_{m''m''},  \\
\label{eq:Fm^i_def}
\bm{\mathcal{F}}^i_{mm'} &=& U'\mob^i_{m'm} + J_{\rm d}\mob^i_{mm'} 
\\
\nonumber
&&
+ \delta_{mm'}J_{\rm H}\sum_{m''}\mob^i_{m''m''},
\end{eqnarray}
and analogous expressions hold for uniform and staggered components of $\mathcal{F}^{(0)}$ and $\bm{\mathcal{F}}$: this corresponds to the replacement $i\rightarrow \rm u$ and $\rm s$ in Eqs.~\eqref{eq:Fn^i_def} and \eqref{eq:Fm^i_def} correspondingly. 
The detailed derivation of Eqs.~\eqref{eq:Fn^i_def} and \eqref{eq:Fm^i_def} is given in Appendix~B. 

%we get the expression for an order parameter in the explicit form
%we express %the order parameter 
From Eqs.~\eqref{eq:S_mm'_def},
(\ref{eq:n_mm_def}--\ref{eq:AFM_general}), see also Eq.~\eqref{eq:order_k_form_alpha} in Appendix~B,  we get the system of mean-field equations in terms of 4-vectors $N^{\rm u}_{mm'}\left(\no^{\rm u}_{mm'}, \mo^{\rm u}_{mm'}\right)$ and $N^{\rm s}_{mm'}\left(\no^{\rm s}_{mm'}, \mo^{\rm s}_{mm'}\right)$
%\begin{eqnarray}
%\label{eq:cq0}
%C'_{m\sigma;m'\sigma'} &=& \frac1{N}\sum'_{\mathbf{k}\alpha}F_{m\alpha\sigma;m'\alpha\sigma'}(\mathbf{k}),\\
%\label{eq:сqQ}
%C''_{m\sigma;m'\sigma'} &=& \frac1{N}\sum'_{\mathbf{k}\alpha}F_{m\alpha\sigma;m'\bar{\alpha}\sigma'}(\mathbf{k}).
%\end{eqnarray}
%From here and from~the 
%we get directly
\begin{eqnarray}
\label{eq:main_HF.n'}
N^{(o)\rm u}_{mm'} &=& \frac1{2N}\sum'_{\mathbf{k}\alpha\sigma}\sigma^{o}_{\sigma\sigma'}F_{m\alpha\sigma;m'\alpha\sigma'}(\mathbf{k}),\\
\label{eq:main_HF.n''}
N^{(o)\rm s}_{mm'} &=& \frac1{2N}\sum'_{\mathbf{k}\alpha\sigma}\sigma^{o}_{\sigma\sigma'}F_{m\alpha\sigma;m'\bar{\alpha}\sigma'}(\mathbf{k}),
\end{eqnarray}
where $o = 0,1,2,3$; a prime over sums denotes summation over the magnetic Brillouin zone and
\begin{equation}\label{eq:F_def}
F_{m\alpha\sigma;m'\alpha'\sigma'}(\mathbf{k}) = \sum_{\nu}{a}^*_{m\alpha\sigma;\nu}(\mathbf{k})a_{m'\alpha'\sigma';\nu}(\mathbf{k})f(E_\nu(\mathbf{k})),
\end{equation}
where $a_{m\alpha\sigma;\nu}$ are eigenvectors belonging to eigenvalue $E_\nu$ of the matrix of the single-electron version of Hamiltonian \eqref{eq:H_MF_total}, see Eq.~\eqref{eq:H_matrix} in Appendix~B, $f(E) = (\exp[(E - E_{\rm F})/T] + 1)^{-1}$ is the Fermi function, and $E_{\rm F}$ is the Fermi level. Within our HFA approach,  the latter is adjusted to set the filling equal to one electron per site. 

We get the expression for the full energy within HFA, $\mathcal{E} = \langle \mathcal{H}_{\rm MF}\rangle$,
\begin{equation}\label{eq:full_energy}
\mathcal{E}/N = \frac1{N}\sum_{\mathbf{k}\nu}E_\nu(\mathbf{k})f(E_\nu(\mathbf{k})) - E_{\rm DC}/N,
\end{equation}
where residual terms of HFA are absorbed into $E_{\rm DC}$, see Eq.~\eqref{eq:E_DC_final} in Appendix~B.
%are introduced, see Appendix~B.

\begin{figure}[t!]
   \centering
  \includegraphics[width=0.5\textwidth]{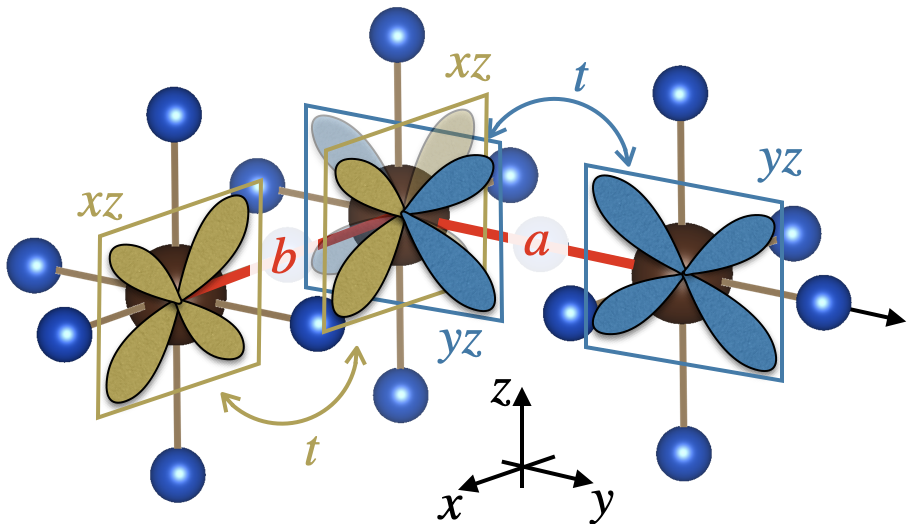}
  \caption{\label{Fig:struct} Sketch illustrating overlap of $xz/yz$ orbitals in the crystal structure of layered perovskites. Transition metals are shown by brown balls, while ligands are  blue. There is an overlap between the $yz$ orbitals in one of the directions ($a$). This orbital stays silent in another direction ($b$), while the $xz$ orbitals strongly overlap. There is also overlap between the $xy$ orbitals in both directions (not shown).}
\end{figure}
The above-presented formalism can be applied in the case of an arbitrary filling, not only 1/6 occupation~(one electron per site). The filling is controlled only by $E_{\rm F}$.  However, the quality of an approximation  depends dramatically on whether  integer or noninteger filling per site situation is considered. In the latter case, for large $U$, strong electron correlations renormalize the electron spectrum, so that more elaborated approximations like dynamical mean-field theory (DMFT) or the Kotliar-Ruckenstein slave-boson approximation~\cite{1986:Kotliar}~(see also calculations in~Refs.~\cite{2015:Igoshev,2018:Igoshev}) should be employed.  In the case of integer filling, HFA is a reasonable approximation.

Finally, we turn to choice of specific parameters. There is an intrinsic deformation --- elongation of transition metal octahedra in layered perovskites, which are physical realizations of a square lattice. This results in such a splitting of the $t_{2g}$ shell that the $xy$ orbital goes higher in energy (by $\Delta_{\rm CF}$) than the $xz/yz$ doublet; see Fig.~\ref{Fig:levels}(a). Another structural feature characteristic for this class of materials is symmetry of hopping parameters. There is always hopping between the  $xy$ orbitals on nearest-neighbor sites $t_{xy/xy} = t$, but the $yz$ orbitals overlap (directly or via the $p_z$ orbital of a ligand) with its partner only along one of the directions ($a$ bond in Fig.~\ref{Fig:struct}), so that $t^{a}_{yz/yz} = t$ and $t^{a}_{xz/xz}=0$. The electrons on the last orbital, $xz$, can also hop only along half of metal--metal bonds ($b$ bond in Fig.~\ref{Fig:struct}): $t^{b}_{xz/xz} = t$ and $t^{b}_{yz/yz} = 0$. Thus, the explicit expressions for the band dispersion reads as
$\varepsilon_{xy,xy}(\mathbf{k}) = -2t(\cos k_x + \cos k_y)$, $\varepsilon_{xz,xz}(\mathbf{k}) = -2t \cos k_x $, $\varepsilon_{yz,yz}(\mathbf{k}) = -2t \cos k_y$. 

Next-nearest-neighbor hopping $t'$ can result in breaking the nesting condition and prevent a metal-insulator transition~\cite{katsnelson1984metal,2023:Igoshev_MIT}, provided that $(U-3J_{\rm H})/t$ is sufficiently small. 
But this case is beyond our consideration. 

%%%%%%%%%%%%%%%%%%%%%%%%%%%%%%%%%%%%%%
\section{Possible states and phase diagram: no spin-orbit coupling}

We start from a somewhat simplified consideration assuming that electrons are localized on particular cubic harmonics, and then go on taking into account quantum effects which result in more complex states obtained by the Hartree-Fock method.

\begin{figure}[t]
   \centering
  \includegraphics[width=0.5\textwidth]{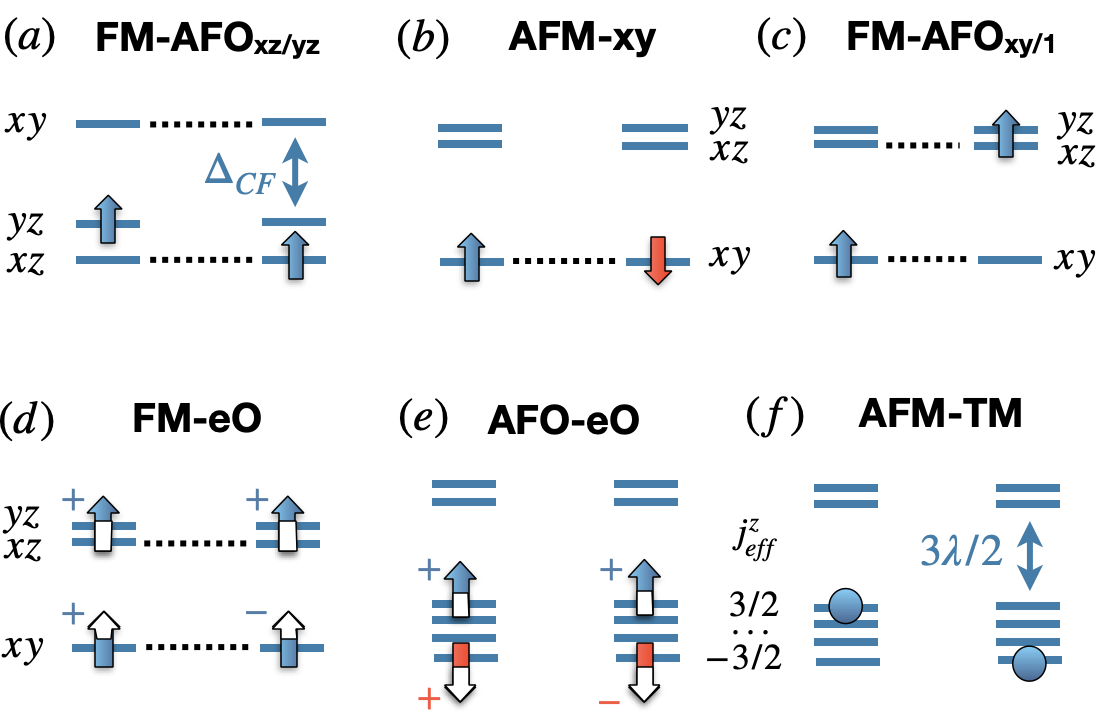}
  \caption{\label{Fig:conf} Sketch illustrating various states for a pair of neighboring sites; corresponding charge-density plots are presented in Fig.~\ref{Fig:orbitals}. Possible hopping paths for (a)--(d) are shown by dashed lines. In the \texttt{FM-AFO$_{xz/yz}$} (ferromagnetic -- antiferro-orbital) configuration, one electron is at the $yz$ orbital, while another one resides at $xz$, and hopping from site to site favors FM spin ordering. Very similar is the situation for \texttt{FM-AFO$_{xy/1}$} ($xy$ on one site and $|l_{\rm eff}^z|=1$ on another are occupied). In the \texttt{AFM}-$xy$ state hopping between the $xy$ orbitals stabilizes AFM order. The other two states, \texttt{FM-eO} and \texttt{AFO-eO}, are orbital-entangled states. The exact wave functions for them are given in \eqref{FM-eO-WF} and \eqref{AFO-eO-WF}. We sketched these wave functions via noncomplete filling of arrows denoting spins; plus and minus signs are used to show the phase of wave functions. In the last \texttt{AFM-TM} state electrons occupy $j^z_{\rm eff} = 3/2$ or $j^z_{\rm eff} = -3/2$ depending on the sublattice (\texttt{TM} stands for the total moment).
  }
\end{figure}

\begin{figure}[t]
   \centering
  \includegraphics[width=0.5\textwidth]{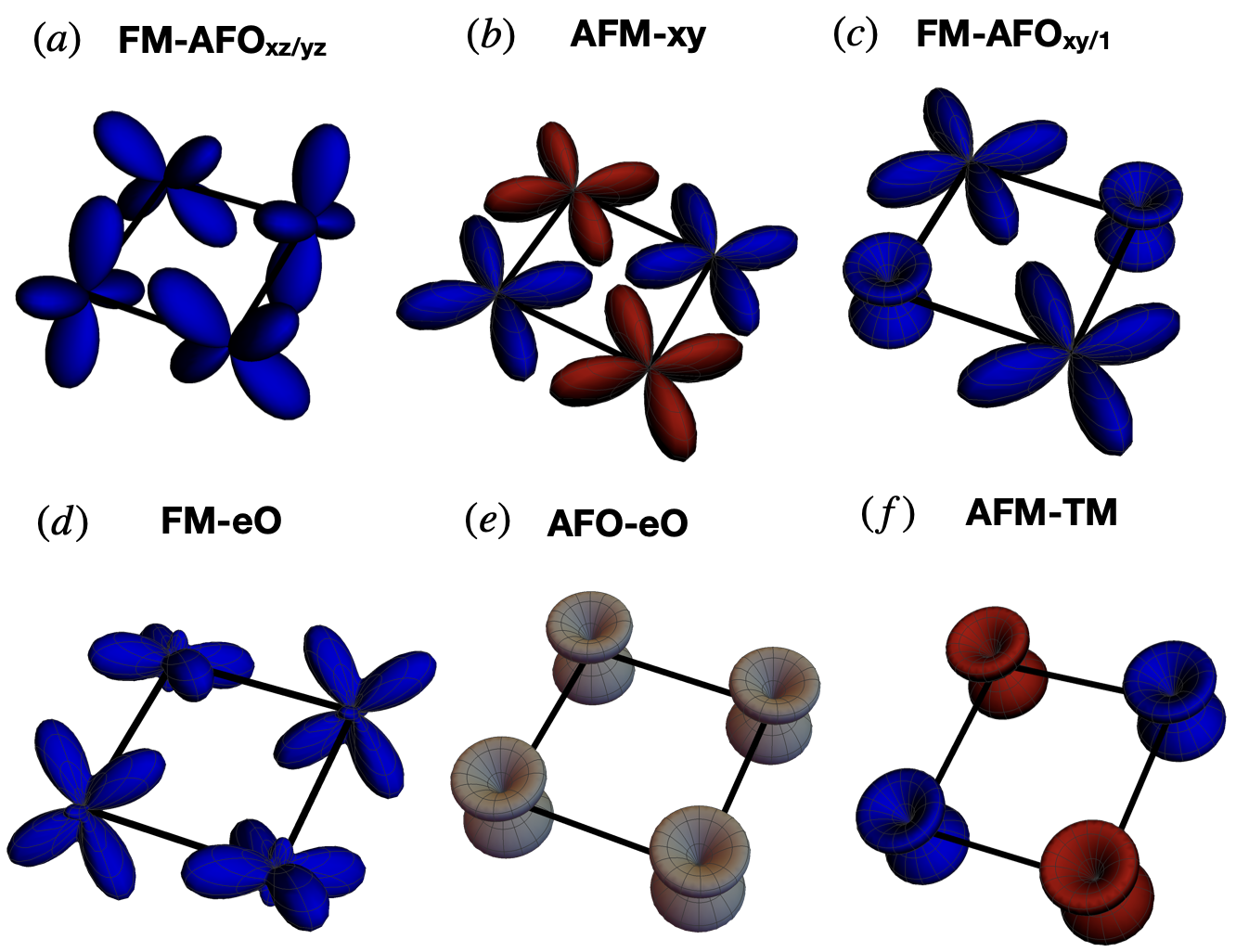}
  \caption{\label{Fig:orbitals} Charge densities for various electronic states under consideration; see Fig.~\ref{Fig:conf}. Two different spins are shown by blue and red. Gray color is used for \texttt{AFO-eO}, where the spin moment vanishes.
  }
\end{figure}

{\it Conventional ferro- and antiferro-orbital states.} 
Without the spin-orbit coupling and for a large and positive crystal-field splitting (i.e., with elongated metal-ligand octahedra), one might expect that in the ground state sites the half-filled $xz$ and $yz$ orbitals will alternate. There is a hopping between half-filled and empty orbitals [antiferro-orbital (AFO) order] for all bonds in this state. This hopping favors ferromagnetic coupling according to Goodenough-Kanamori-Anderson (GKA) rules~\cite{Goodenough,khomskii2014transition,Khomskii2021}. Such a state is shown in Fig.~\ref{Fig:conf}(a) and Fig.~\ref{Fig:orbitals}(a) and referred to as \texttt{FM-AFO$_{xz/yz}$} in what follows. This type of orbital and magnetic order is favored by  Hund's intra-atomic exchange $J_{\rm H}$ and large crystal-field field $\Delta_{\rm CF}$~\cite{Khomskii2021} and realized, e.g., in perovskite YTiO$_3$~\cite{2005:Streltsov_Mylnikova}. 

An alternative  \texttt{AFM}-$xy$ state is shown in Fig.~\ref{Fig:conf}(b) and Fig.~\ref{Fig:orbitals}(b). Electrons occupy the $xy$ orbital at all sites in this case. This ferro-orbital (FO) ordering is stabilized by a small positive or negative $\Delta_{\rm CF}$ and, according to GKA rules, leads to the antiferromagnetic (AFM) state. 

\begin{figure}[t!]
   \centering
  \includegraphics[angle=-90,width=0.5\textwidth]{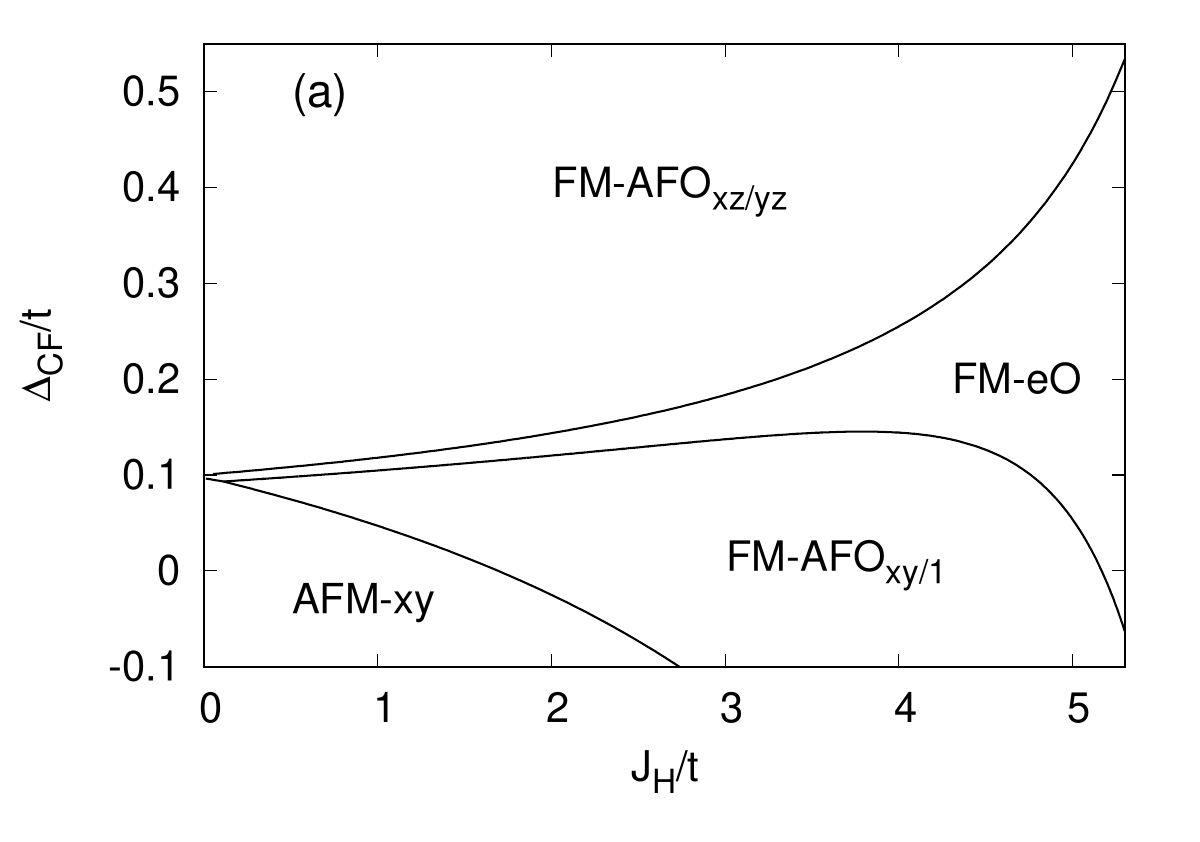}
  \includegraphics[angle=-90,width=0.5\textwidth]{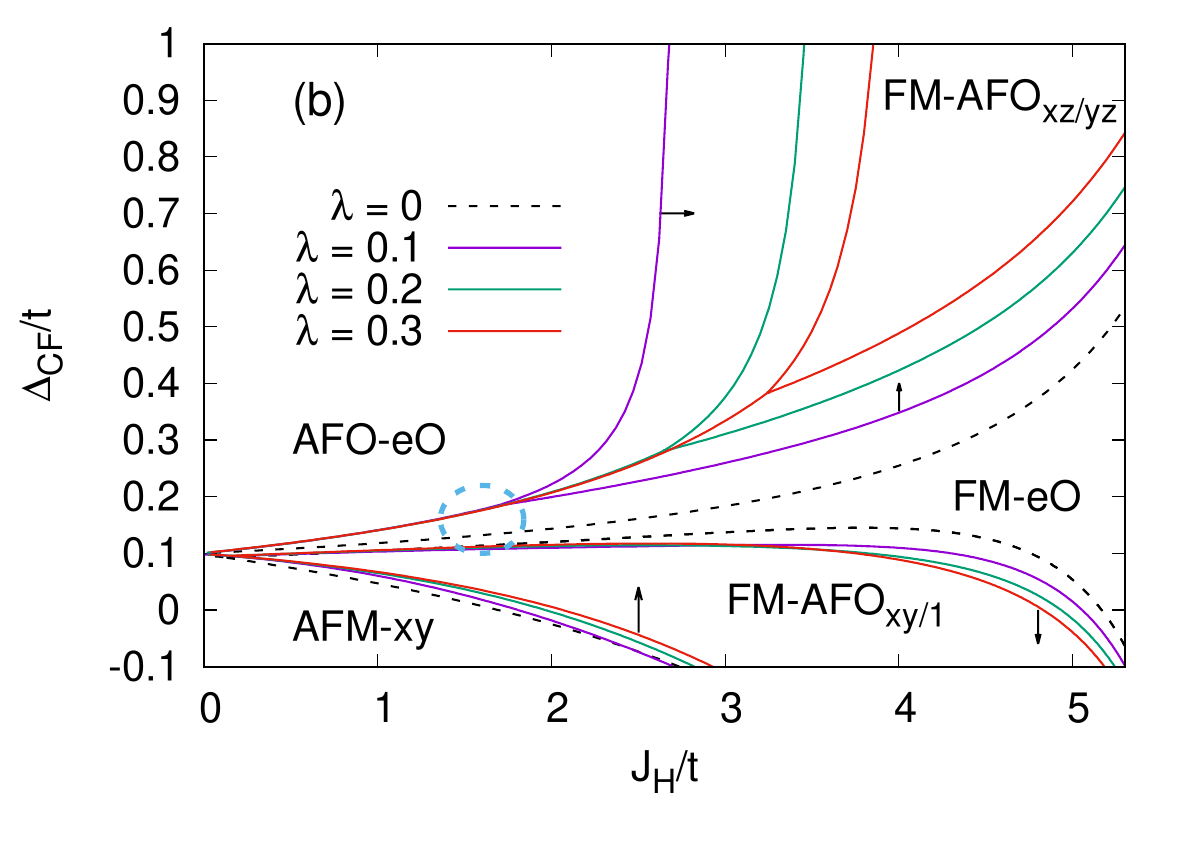}
  \caption{\label{Fig:PD-lambda0} Ground state phase diagram for the crystal-field splitting $-0.1t<\Delta_{\rm CF}< 1.0t$  without (a) and  taking into account (b) the spin-orbit coupling. Here $J_{\rm H}$ stands for the Hund's intra-atomic exchange, $\lambda$ is the spin-orbit coupling constant, $t$ is the hopping, and Hubbard's parameter is $U=20t$. 
  Sketches of different states are shown in Figs.~\ref{Fig:conf} and \ref{Fig:orbitals}. 
  An increase of $\lambda$ results in an expansion of \texttt{AFO-eO} phase region to the right, an expansion of \texttt{FM-eO} phase region upward, and contraction of \texttt{FM-AFO$_{xz/yz}$} (the directions of phase boundaries moving are shown by arrows). In (b) black dotted line shows the result for $\lambda = 0$ [the same as in Fig.~\ref{Fig:PD-lambda0}(a)] for the sake of convenience. Blue circle indicates the region of relevance for Sr$_2$VO$_4$ parameters (close to $J_{\rm H} = 1.6t$, $\Delta_{\rm CF} = 0.16t$); see Sec.~\ref{Sec:Sr2VO4}.  
}
\end{figure}

The full phase diagram including these two states obtained by the Hartree-Fock method is presented in Fig.~\ref{Fig:PD-lambda0}(a). It was calculated by direct solution of the nonlinear self-consistent system (independent 71 variables) of Eqs.~(\ref{eq:main_HF.n'} and \ref{eq:main_HF.n''}) comparing full energies of different phases in the $T\rightarrow 0$ limit, see~Eqs.~(\ref{eq:full_energy}) and~(\ref{eq:E_DC_final}). The self-consistency process starts with choice of initial states and iterative procedure modifying them. Brillouin zone integration in $\mathbf{k}-$space  was performed by triangular method~\cite{2002:BZ_integration:Joo-Hyoung} for $N_{\rm grid} = 40$ triangles. 

Negative crystal-field splitting $\Delta_{\rm CF}$ (when the $xy$ orbital gets lower than the $xz/yz$ doublet) obviously stabilizes the \texttt{AFM}-$xy$ state, but it is realized even for small and positive $\Delta_{\rm CF}$. This is because for \texttt{AFM}-$xy$ both electrons tunnel and lower the total energy for any bond, while for \texttt{FM-AFO$_{xz/yz}$} only one electron can hop along each bond ($yz$ for bond $a$ and $xz$ for bond $b$). On the other hand, one can see from Fig.~\ref{Fig:PD-lambda0} that the \texttt{AFM}-$xy$ state is destabilized by the Hund's coupling. In this case the energy of the excited state $E_{\rm exc}$ (due the hopping) is strongly reduced by $J_{\rm H}$ in the case of \texttt{FM-AFO$_{xz/yz}$} (one electron is on $xz$, while another one is on the $yz$ orbital; $E_{\rm exc}=U-3J_{\rm H}$, if we use the Kanamori parametrization~\cite{Kanamori1963}) with respect to what we have for \texttt{AFM}-$xy$ (both electrons are on the $xy$ orbital; $E_{\rm exc}=U$). 

However, strong Hund’s coupling stabilizes not \texttt{FM-AFO$_{xz/yz}$}, but a very different state, with electrons occupying the $xy$ orbital on sublattice~A and an arbitrary superposition of the $xz$ and $yz$ orbitals with the same spin projection on the B sublattice. For example, one can chose the following wave function,
\begin{eqnarray}
\label{lz=pm1}
| l^z_{\rm eff}=1 \rangle = - \frac 1{\sqrt 2}(| yz\rangle +  \I |xz \rangle)
\end{eqnarray}
see in Fig.~\ref{Fig:conf}(c) and Fig.~\ref{Fig:orbitals}(c). We note that any other mixture of the $xz$ and $yz$ orbitals can be used without the spin-orbit coupling. In this \texttt{FM-AFO$_{xy/1}$} state we win both by intra-atomic exchange (in the excited state both electrons have the same spin) and by a very efficient hopping of electrons on the $xy$ orbitals.  This state is doubly degenerate with respect to spin [if we take into account the spin-orbit coupling the spin-down electrons occupy the $|l_{\rm eff}^z=-1 \rangle = (| yz\rangle -  \I |xz \rangle)/ \sqrt 2$ orbital on sublattice B].

{\it \texttt{FM-eO} phase with entangled orbitals (\texttt{eO}).}
However, hopping can be optimized even further on mixing $xy$ and $yz/xz$ orbitals (or their linear combinations given by $l_{\rm eff}^z=\pm 1$) in an appropriate way. 
%There is another state in the phase diagram shown in Fig.~\ref{Fig:PD-lambda0}(a).  As it has been mentioned above, neither \texttt{AFM}-$xy$ nor \texttt{FM-AFO} are optimal with respect to exchange processes via the $xy$ or $yz/xz$ orbital channels simultaneously. Mixing them in an appropriate way we considerably reduce the total energy, and a new phase stabilized by quantum entanglement of orbitals (eO) appears for intermediate set of parameters, see Fig.~\ref{Fig:PD-lambda0}. 
This new phase is characterized by FM coupling and referred to as \texttt{FM-eO}. In the case of degenerate $t_{2g}$ orbitals ($\Delta_{\rm CF}=0$) one can find the wave functions for two sites forming a checkerboard order analytically~\cite{2023:JMMM:Igoshev_Streltsov_Kugel}: 
\begin{eqnarray}
\label{FM-eO-WF}
\text{A}: \quad \frac 23  |xy ,\sigma \rangle +  \frac{\sqrt {5}}3 | \ l^z_{\rm eff}=1, \sigma \rangle,\\
\text{B}: \quad \frac 23|xy, \sigma \rangle -  \frac{\sqrt {5}}3 | \ l^z_{\rm eff}=1, \sigma \rangle, \nonumber
\end{eqnarray}
where A and B are two sublattices, and $\sigma$ stands for spin. 
It has to be mentioned that this solution  is degenerate  with respect to $\sigma \to -\sigma$ and $l^z_{\rm eff} \to -l^z_{\rm eff}$ inversions separately. However, for a finite spin-orbit coupling, see below, only the symmetry $(l^z_{\rm eff},\sigma) \to (-l^z_{ \rm eff},-\sigma)$ remains. Second, it is remarkable that two sublattices differ only by the phase of $|\ l^z_{\rm eff}=1, \sigma \rangle$ component. Finally, in a general case of arbitrary $\Delta_{\rm CF}$ exact expression for wave functions depends on parameters and changes in different points of the phase diagram. This is in a contrast to the \texttt{AFO-eO} phase discussed in Sec.~\ref{Sec:PD-SOC}.

At $J_{\rm H} = 0$ only \texttt{AFM}-$xy$ and \texttt{FM-AFO$_{xz/yz}$} are presented: the critical point is determined by competition of (intersite) exchange  $J_{\rm ex} = t^2/U$ and the crystal-field splitting $\Delta_{\rm CF}$: $\Delta^{*}_{\rm CF} = 2J_{\rm ex}$. Stability of the \texttt{FM-eO} and \texttt{FM-AFO$_{xy/1}$} phases is rapidly increased and corresponding regions on the phase diagram expand as $J_{\rm H}$ increases. Parameters corresponding to Sr$_2$VO$_4$ at normal conditions are presented Sec.~\ref{Sec:Sr2VO4}. They are rather close to the phase boundary between the \texttt{FM-AFO$_{xz/yz}$}, \texttt{FM-AFO$_{xy/1}$}, and \texttt{FM-eO} states, but the phase diagram by itself strongly changes by the spin-orbit coupling as we show in the next section.

%%%%%%%%%%%%%%%%%%%%%%%%%%%%%%%%%%%%%%
\section{Phase diagram at finite spin-orbit coupling \label{Sec:PD-SOC}}
{\it \texttt{AFO-eO} phase stabilized by SOC.} Generally speaking, the crystal-field splitting and the spin-orbit coupling (SOC) tend to stabilize electrons on quite different orbitals. The first one prefers real (cubic) wave functions, while the second prefers complex spherical harmonics (this can be critical, in particular, for the Jahn-Teller effect~\cite{StreltsovPRX,streltsov2022prx}). The intra-atomic exchange $J_{\rm H}$ favors a maximal total spin (in our situation this is important for virtual excited states with two electrons per site), whereas wave functions stabilized by SOC can mix different spin components; see, e.g., Eqs.~(28--29) in~Ref.~\cite{Khomskii2021}. Therefore, SOC must affect the phase diagram of the three-orbital Hubbard model described above. 

First of all, SOC is expected to influence the states with partially filled degenerate $xz/yz$ orbitals, i.e. the situation shown in Fig.~\ref{Fig:levels}(a) corresponding $\Delta_{\rm CF}>0$. Indeed, one can always gain some energy due to the SOC putting our electron on a linear combination of these orbitals corresponding to $l^z_{\rm eff}=1$; see Eq.~\eqref{lz=pm1}. This is shown in Fig.~\ref{Fig:levels}(c), where the $j_{\rm eff}=3/2$ quartet is split by a positive crystal field on two Kramers doublets: $j_{\rm eff}^z=\pm 1/2$ and $j_{\rm eff}^z=\pm 3/2$. The results of direct Hartree-Fock calculations presented in Fig.~\ref{Fig:PD-lambda0}(b) demonstrate that even a modest SOC with $\lambda = 0.1t$ leads to shift of the \texttt{FM-AFO$_{xz/yz}$} state to the region of large $J_{\rm H}$ and to formation of a novel \texttt{AFO-eO} phase, sketched in Fig.~\ref{Fig:conf}(e) and Fig.~\ref{Fig:orbitals}(e). The wave function in this new phase is written in a very simple form:
\begin{eqnarray}
\label{AFO-eO-WF}
\frac 1{\sqrt{2}} \Big( 
| \ l^z_{\rm eff}=1, \uparrow \rangle
\pm | \ l^z_{\rm eff}=-1, \downarrow \rangle
\Big),
\end{eqnarray}
where different signs are taken for two different sublattices (A or B). This \texttt{AFO-eO} state is exactly the state with staggered order of in-plane confined isospins, found by Jackelli and Khaliullin in \cite{Jackeli2009a}. Dipole spin 
\begin{equation}\label{S-def}
\mathbf{S}_i = \sum_m\bm{\mathcal{S}}_{imm}
\end{equation}
[with generalized spin operator $\bm{\mathcal{S}}_{imm}$ defined in \eqref{eq:S_mm'_def}] and orbital $\mathbf{L}_i$ moments vanish at every lattice site  $\langle  \mathbf{S}_i \rangle =\langle   \mathbf{L}_i \rangle = 0$, so that only octupole moments remain non-zero and they order in a staggered fashion, as we will discuss below.

{\it Higher-lying in energy \texttt{AFM-TM} phase.}
A very different state was considered in \cite{Eremin2011}. In particular, the electron can be localized not on a linear combination of $|j_{\rm eff}^z = +3/2 \rangle $ and $|j_{\rm eff}^z = -3/2 \rangle$ at the given sublattice as in the case of the \texttt{AFO-eO} state, but it can be described by ``pure'' $\psi_{\rm A} = |j_{\rm eff}^z = +3/2 \rangle$ at the A and $\psi_{\rm B} = |j_{\rm eff}^z = -3/2 \rangle$ at the B sublattice. The latter situation is shown in Figs.~\ref{Fig:conf}(f) and ~\ref{Fig:orbitals}(f) and dubbed as \texttt{AFM-TM}  (total angular momentum antiferromagnetism). Spin and orbital momenta exist on each site in this case, but cancel each other, as explained in \cite{Eremin2011}.

We performed an accurate consideration of the energies of both \texttt{AFO-eO} and \texttt{AFM-TM} states within the mean-field approximation for the kinetic exchange Hamiltonian derived from~Eq.~(\ref{eq:hamilt}) in the limit $zt\ll U - 3J_{\rm H}$ using the formalism developed in \cite{2023:JMMM:Igoshev_Streltsov_Kugel} ($z$ is the coordination number; see Appendix~A for details). 
While both these states are stabilized mainly by spin-orbit coupling, in the leading (zeroth) order with respect to $\lambda$ and $\Delta_{\rm CF}$ the  kinetic exchange Hamiltonian yields that the \texttt{AFO-eO} state has a lower energy than the \texttt{AFM-TM} state due to the explicitly nonsymmetric (with respect to $i\leftrightarrow j$) \textit{ferromagnetic} exchange contribution in the effective Hamiltonian 
\begin{multline}\label{eq:H_kin_non-sym_exchange_term}
\mathcal{H}_{\rm non-symm} = -\frac{2J_{\rm H}}{(U - 3J_{\rm H})(U - J_{\rm H})}\\
\sum_{ijm}|t^m_{ij}|^2\bm{\mathcal{S}}_{imm}\mathbf{S}_j.
\end{multline}
See the derivation in Appendix~B. 
It can be readily shown that for the $\texttt{AFM-TM}$ state this term is strictly positive whereas for the $\texttt{AFO-eO}$ state it vanishes  (all other terms in the kinetic exchange Hamiltonian yield the same result for both states). This analytic result can be compared with direct Hartree-Fock calculations presented in Fig.~\ref{Fig:AFMoe-TAFM} in Appendix~A. Both approaches clearly show that \texttt{AFM-TM} is always higher in energy than \texttt{AFM-TM} and for small $J_{\rm H}$ this difference is well described by \eqref{eq:H_kin_non-sym_exchange_term}.

%%%%%%%%%%%%%%%%%%%%%%%%%%%%%%%%%%%%%%%%%%%%%%%%
%% Finite lambda phase diagram:
%%%%%%%%%%%%%%%%%%%%%%%%%%%%%%%%%%%%%%%%%%%%%%%%
{\it Competition of \texttt{AFO-eO}/\texttt{FM-eO} and \texttt{FM-AFO$_{xz/yz}$} phases.}
Coming back to the \texttt{AFO-eO} phase one can notice that  SOC is very efficient in suppression  of the standard \texttt{FM-AFO$_{xz/yz}$} state even at relatively large $J_{\rm H}$; see Fig.~\ref{Fig:PD-lambda0}(b). There is a very similar effect on narrowing the phase region of \texttt{FM-AFO$_{xz/yz}$} state  and expansion of not only the \texttt{AFO-eO} phase, but the \texttt{FM-eO} solution with increasing SOC as well.

\begin{figure}[t]
   \centering  \includegraphics[angle=-90,width=0.5\textwidth]{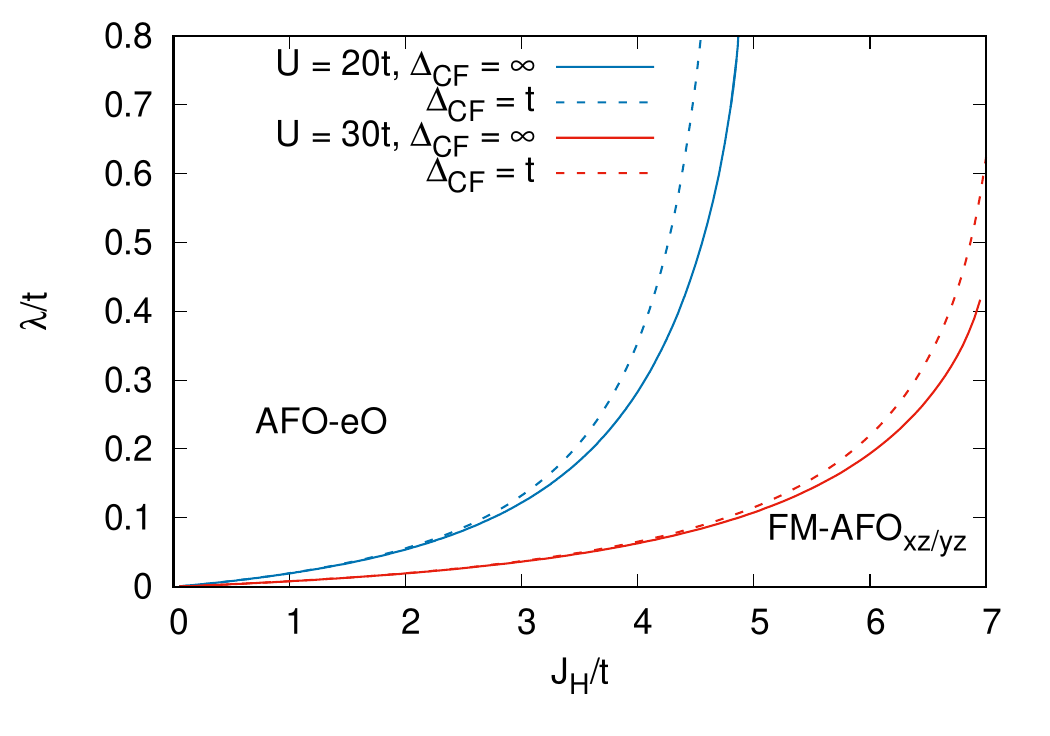}  \caption{\label{Fig:AFMoe-FMAFO}
Ground state phase diagram for large ($\Delta_{\rm CF} = 1.0t$; dashed lines) or infinite ($\Delta_{\rm CF} = \infty$; solid lines) in terms of $J_{\rm H}$ and $\lambda$. Blue (red) lines show the $U = 20t~(30t)$ case.  Two phases, \texttt{FM-AFO}$_{xz/yz}$ and \texttt{AFO-eO}, are considered.  
}
\end{figure}

%%%%%%%%%%%%%%%%%%%%%%%%%%%%%%%%%%%%%%%%%%%%%%%%
%% Large Delta_CF phase diagram:
%%%%%%%%%%%%%%%%%%%%%%%%%%%%%%%%%%%%%%%%%%%%%%%%
For comparison, in~Fig.~\ref{Fig:AFMoe-FMAFO} we present the phase diagram in other ($J_{\rm H} - \lambda$) variables for a typically (realized experimentally) situation of positive and large $\Delta_{\rm CF}$. Such a crystal field shifts the $xy$ orbital upward and this destabilizes the \texttt{AFM}-$xy$ and \texttt{FM-eO} phases [where electrons occupy the $xy$ orbital; see Figs.~\ref{Fig:conf}(b) and (d)], so that only the \texttt{AFO-eO} and \texttt{FM-AFO$_{xz/yz}$} states survive. At small and moderate Hund's exchange, the \texttt{AFO-eO} solution dominates being stabilized by finite $\lambda$. In contrast, large $J_{\rm H}$ favors the \texttt{FM-AFO$_{xz/yz}$} state. Indeed, excited (intermediate) states in the exchange processes for the \texttt{FM-AFO$_{xz/yz}$} solution obey Hund's rule (these are triplet states like $xz_\uparrow yz_\uparrow$). Excited states for the \texttt{AFO-eO}  phase do not completely optimize intra-atomic exchange interaction.

The lowering of $\Delta_{\rm CF}$ favors the \texttt{FM-AFO$_{xz/yz}$} phase which changes the phase diagram by expansion of the corresponding phase region quantitatively, but not qualitatively. However the properties of the latter state strongly depend on parameter values, as explained in the next section. Last but not least, there is a strong influence of Hubbard $U$ on the phase diagram, which is clearly seen in Fig.~\ref{Fig:AFMoe-FMAFO}.

%%%%%%%%%%%%%%%%%%%%%%%%%%%%%%%%%%%%%%%%%%%%%%%%
%% Symmetry of AFM-oe state
%%%%%%%%%%%%%%%%%%%%%%%%%%%%%%%%%%%%%%%%%%%%%%%%

{\it Order parameters.}
While the \texttt{FM-AFO$_{xz/yz}$}, \texttt{FM-AFO$_{xy/1}$}, \texttt{AFM-TM}, and \texttt{FM-eO} states can be characterized by standard order parameters --- dipole moments $\langle  \mathbf{L}_i \rangle$ and $\langle \mathbf{S}_j \rangle$ on corresponding sublattices $i=\{A,B\}$ --- in \texttt{AFM-eO} both such parameters vanish, as we have discussed above (hidden orbital-spin order~\cite{Jackeli2009a}).

The \texttt{AFM-eO} state is stabilized by strong SOC putting electrons on the spin orbitals characterized by the total angular moment $\mathbf{J}$ (in our case $\mathbf{j}_{\rm eff}$). It is instructive to consider expectation values of not only dipole (transforming according to $\Gamma_4$ representation of cubic group), but also quadrupole ($\Gamma_3$ and $\Gamma_5$ representations) and octupole  ($\Gamma_2$, $\Gamma_4$, and $\Gamma_5$ representations) moments, which can be expressed via corresponding combinations of $\mathbf{J}$ components ~\cite{Santini2009}.  

Our situation is characterized by the tetragonal symmetry in which $\Gamma_4$ and $\Gamma_5$ representations are reducible and direct calculations show that only two out of seven octupoles,  $\langle T^\alpha_x\rangle$ and $\langle T^\beta_x\rangle $, transforming over corresponding one-dimensional irreducible representations, are nonzero for the \texttt{AFM-eO}~state (all dipoles and quadrupoles vanish). There is antiferro-octupole ordering for both of them (nearest neighbors have opposite octupole moments). Moreover, on the same lattice site   $\langle T^\alpha_x\rangle$  and $\langle T^\beta_x\rangle$ have the same sign. 
%\sout{This situation reminds the intra-atomic Hund's exchange, which makes  spins  parallel.} 

%%%%%%%%%%%%%%%%%%%%%% 
\section{Modification of the Kugel-Khomskii FM-AFO$_{xz/yz}$ order by SOC}

As shown in the previous section, while SOC generates new states having different anomalous properties such as vanishing dipole magnetic moment, it also modifies conventional states stabilized by the kinetic exchange. In this section we consider the evolution of the \texttt{FM-AFO$_{xz/yz}$} state with increasing SOC. 

\begin{figure}[t]
   \centering  \includegraphics[angle=-90,width=0.5\textwidth]{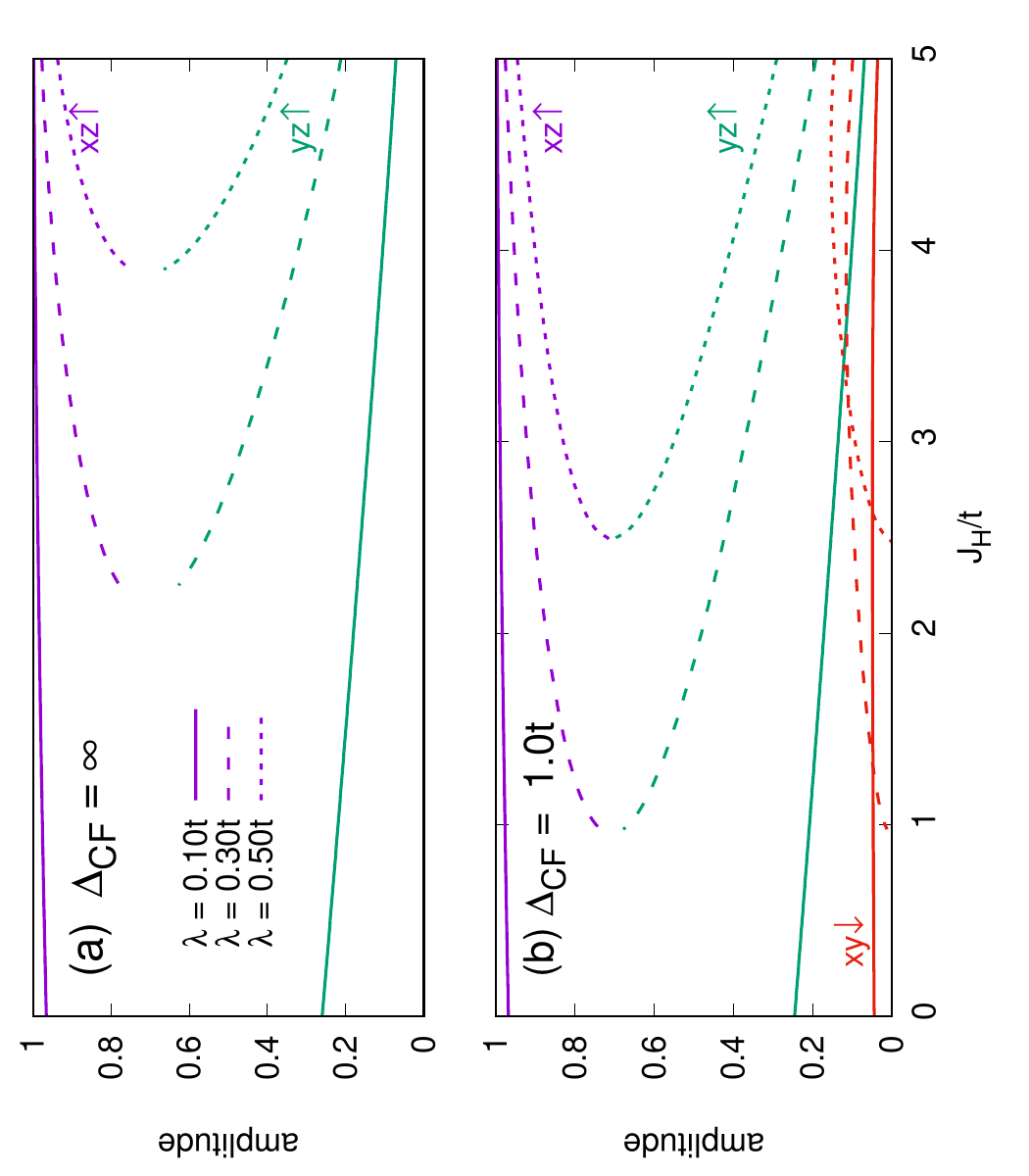}  \caption{\label{Fig:WF}
Absolute value of  wave function amplitudes of the \texttt{FM-AFO}$_{xz/yz}$~state 
for one of the sublattices depending on $J_{\rm H}$ at $\Delta_{\rm CF} = \infty$ (a) and $\Delta_{\rm CF} = 1.0t$~(b), where $t$ is the hopping.  Solid lines correspond to $\lambda = 0.1t$, dashed lines to $\lambda = 0.3t$, dotted lines to $\lambda = 0.5t$. Contribution of $xz\uparrow$ states is shown in violet, $yz\uparrow$ in green, and $xy\downarrow$ in red.
}
\end{figure}

In Fig.~\ref{Fig:WF} we present decomposition of the occupied state (obtained by diagonalization of the on-site occupation matrix for one of sublattices) in cubic harmonics. In the case of  very large crystal field, $\Delta_{\rm CF} = \infty$, presented in the upper panel of Fig.~\ref{Fig:WF}, we have a conventional \texttt{FM-AFO$_{xz/yz}$} with the electron sitting mostly at the $xz \uparrow$ orbital (for the second sublattice it will be $yz \uparrow$). The intra-atomic Hund's exchange works for this state due to the gradually reducing contribution of other orbitals caused by a finite SOC, $\lambda=0.1t$.  

However, a further increase of $\lambda$ (up to $0.3t$) changes the situation dramatically. The SOC is nearly incapable of struggling with a too large crystal field and therefore faintly affects the $xy$ orbital. In contrast, it mixes the $xz$ and $yz$ orbitals to form $l^{z}_{\rm eff} = \pm 1$ states with the same spin projection, \eqref{lz=pm1}, which are eigenfunctions of SOC. This allows a gain in energy as much as possible, both in the crystal field and  SOC contributions, but breaks the first Hund's rule. As a result, for fixed $\lambda > 0$ there exists a critical $J_{\rm H}^*$ below which occupations of $xz\uparrow$ and $yz\uparrow$ coincide [coefficients at corresponding wave functions are close to $1/\sqrt2$, see Fig.~\ref{Fig:WF}(a), while orbital occupancies equal 0.5]. The price to pay is a reduced energy gain due to kinetic exchange.  For a finite crystal field, one observes a quite similar tendency of mixing the $xz$ and $yz$ orbitals and onset of critical $J_{\rm H}^*$; see Fig.~\ref{Fig:WF}(b). 
We also find that lowering $\Delta_{\rm CF}$ stabilizes the \texttt{FM-AFO}$_{xz/yz}$ state, increasing the difference in occupation of the $xz \uparrow$ and $yz \uparrow$ orbitals as well as a nonvanishing contribution of the $xy\downarrow$ state.

%%%%%%%%%%%%%%%%%%%%%%%%
\section{Application to S\lowercase{r}$_2$VO$_4$\label{Sec:Sr2VO4}}

As explained in the introduction section, a physical realization of the three-orbital Hubbard model on a square lattice with a single electron is Sr$_2$VO$_4$ with quite unusual magnetic properties, which are still to be understood.  
We performed density function calculations of Sr$_2$VO$_4$ taking experimental crystal structure~\cite{Cyrot1990}. Further details are as follows: Approximation for the exchange-correlation potential was chosen following Perdew, Burke, and Ernzerhof (PBE)~\cite{Perdew1996}, the VASP code~\cite{Kresse1996} was applied, and the projector augmented-wave method \cite{Blochl1994}  together with integration over tetrahedra \cite{Blochl1994a}  with 405 $\mathbf{k}$ points was used;  for projection onto a small Hamiltonian with only $t_{2g}$ orbitals we applied the projected localized orbitals technique~\cite{2018:Schuler}. Some restricted set of orbital orders and collective excitations above them were considered within HFA to describe Sr$_2$VO$_4$ in the three-orbital Hubbard model at particular values of model parameters~\cite{2023:Mohapatra:Sr2VO4}. However, the complete ground state phase diagram and some important states were not considered there. 

The hopping between the $xz/yz$ orbitals was found to be $t=$250 meV (very close to the previous estimate \cite{Arita2012}), while the crystal-field splitting within $t_{2g}$ is $\Delta_{\rm CF}=40$ meV (i.e., $xy$ orbital lies higher than the $xz/yz$ doublet), so  that $\Delta_{\rm CF}/t = 0.16$. The spin-orbit coupling constant for V$^{3+}$ is about 30 meV ($\lambda/t = 0.12$)~\cite{Abragam}. 

The constrained random-phase approximation (cRPA) has been used in Ref.~\cite{Kim2017b} to calculate interaction parameters. For the $t_{2g}$-only model used in the present work they turn out be $U=2.7$ eV and $J_{\rm H}=0.4$ eV ($U=11t$, $J_{\rm H}=1.6t$)~\cite{Kim-Francini}. In other papers, Hubbard's $U$ varies from 3 to 5.3 eV ($U=12-21t$)~\cite{2007:Arita,Kim2017b,moore2024}, depending on how exactly  $U$ is calculated and what is included in screening channels. One can see that with these characteristic parameters Sr$_2$VO$_4$ is situated very close to boundaries of the \texttt{AFO-eO}, \texttt{FM-eO}, \texttt{FM-AFO$_{xz/yz}$}, and \texttt{FM-AFO$_{xy/1}$} phases; see Fig.~\ref{Fig:PD-lambda0}(b). Thus suppressed magnetic moment (or its absence) can be related to formation of orbital entangled  \texttt{AFO-eO} or \texttt{FM-eO} states, depending on the specific parameters realized in the system.

Strictly speaking, the term ``entangled'' does not mean here quantum entanglement, but only reflects quantum superposition at the one-electron level.  Although spin-orbit coupling generally describes the interaction of spin and orbital operators and mixing of corresponding degrees of freedom,  its impact depends strongly on the specific state: for the \texttt{AFO-oe} phase it is the main one, and for other phases, it is secondary.  Besides, for the \texttt{FM-eO} phase there is the phase change of $t_{2g}$ orbital amplitude going from one sublattice to another: two sublattices differ by this sign only. Obviously, such a mixing can be taken into account within HFA.

For the sake of completeness, we also performed optimization of the crystal structure for uniform and not too high pressure of 10 GPa (at very high $P$, the system is expected to become metallic and the physical mechanism responsible for magnetism and other physical properties will change). We found that the crystal field increases to 64 meV; the hopping turns out to be 295 meV and thus also grows to $\Delta_{\rm CF} = 0.22t$. On the one hand, this works for stabilization of \texttt{AFO-eO} states shifting Sr$_2$VO$_4$ higher and to the left ($J_{\rm H}/t$ effectively decreases) in the phase diagram of Fig.~\ref{Fig:PD-lambda0}(b). On the other hand, while in the first approximation atomic-like parameters $J_{\rm H}$ and $\lambda$ are not expected to change considerably, it is not {\textit a priori} clear how strongly pressure will affect screening of $U$.

%%%%%%%%%%%%%%%%%%%%%%%%%%
\section{Conclusions}

In this work we considered interplay between different interactions, which define the ground state properties of the three-orbital model on a square lattice with a single electron per site relevant, e.g., for layered perovskites with $t^1_{\rm 2g}$ electronic configuration. The method used --- the generalized Hartree-Fock approximation --- is suitable to describe the case of strong electronic correlations at other integer filling as well. However,  more elaborated approaches such as dynamical mean field theory~\cite{samani2024map} should be applied in the metallic regime.

We also consider only the checkerboard order and demonstrate that even this very particular case has a rich phase diagram and the spin-orbit coupling changes it dramatically leading to stabilization of several highly unusual states including those where conventional spin moment vanishes and an exotic order of octupolar magnetic moments appears. 

There are five different phases according to present calculations: three conventional states with antiferromagnetic and ferro-orbital, \texttt{AFM}-$xy$, and ferromagnetic and antiferro-orbital orderings, \texttt{FM-AFO$_{xz/yz}$} and \texttt{FM-AFO$_{xy/1}$}, stabilized by the Kugel-Khomskii mechanism and two other orbital-entangled states --- \texttt{AFO-eO} and \texttt{FM-eO}. The ferromagnetic \texttt{FM-eO} phase is favored by the intra-atomic Hund's exchange and competition of different exchange interaction is ferromagnetic. Another unconventional state, \texttt{AFO-eO}, is promoted by the spin-orbit coupling. This phase was proposed by Jackeli and Khaliullin \cite{Jackeli2009a}, while the state suggested by Eremin {\it et al.}~\cite{Eremin2011} turns out to be higher in energy.  

It is worth noting that a plethora of various states in the model under consideration is due to two reasons: (1) competition of different interactions (spin-orbit coupling, inter-site and intra-atomic exchanges) and (2) symmetry of the problem, which makes hopping processes strongly orbital dependent and anisotropic. 

Our consideration can be readily generalized to treat different transition metal compounds with anomalous physical properties related to the spin-orbit coupling and interplay between spin and orbital degrees of freedom, e.g., Ba$_2$NaOsO$_6$~\cite{Erickson2007,mosca2021interplay,Lu2017,mosca2024mott}, Ba$_2$CeIrO$_6$~\cite{revelli2019spin}, or Ba$_2$CaOsO$_6$~\cite{Voleti2020,Pourovskii2021,Thompson2014,2021:Khaliullin}. The results obtained for Sr$_2$VO$_4$ demonstrate that this material lies close to the region of the phase diagram, where four out of five states can be realized. Therefore some external perturbation can be used to change physical properties of Sr$_2$VO$_4$ shifting it from one to another phase.  While most of experimental results have been obtained in previous decades, physical properties of this material are still intriguing and remain unexplained. Present high-precision neutron diffraction experiments with error bar smaller than 0.6$\mu_{\rm B}$ \cite{Cyrot1990} might be useful to resolve the issue with vanishing magnetic moment. In the meantime, other  modern experimental techniques~\cite{Sibille2020,Sasabe2021,lovesey2021} can be used to study possible octupolar order in Sr$_2$VO$_4$.

Theoretically it would be very interesting to include in consideration not only orbital degrees of freedom and the spin-orbit coupling as was done in the present paper, but also to take into account a long-range exchange interaction and its possible anisotropy, since both have been demonstrated to play an important role for Sr$_2$VO$_4$\cite{Kim2017b}.

%%%%%%%%%%%%%%%%%%%%%%%%%%%%%%%%%%%%%%
\section{Acknowledgements}
S.S. thanks G. Khaliullin and K. Kugel for useful discussions and B. Kim and C. Franchini for unpublished estimates of Hubbard's $U$ and Hund's $J_{\rm H}$ using cRPA for different models and for various stimulating communications.

The research funding from the Ministry of Science and Higher Education of the Russian Federation (the state assignment, theme ``Quantum'' No. 122021000038-7) for implementation of generalized Hartree-Fock method in computer codes is acknowledged. The application of this treatment to perovskites is supported by the  Russian Science Foundation, Project No. 23-42-00069.

%\bibliography{library-web-mendeley,igoshev_pa.bib}

\appendix
%%%%%%%%%%%%%%%%%%%%%%%%%%%%%%
\section{Effective model in a multiorbital case} \label{appendix:kinetic_exchange_derivation}

In this appendix we derive the effective kinetic Hamiltonian for arbitrary hopping matrix $t^{mm'}_{ij}$ generalizing the classical derivation of Kugel and Khomskii~\cite{KK-UFN}. 
We consider the Hamiltonian (\ref{eq:hamilt}) in the case of one electron per site, treating $\mathcal{H}_0 = \mathcal{H}_{\rm CF} + \mathcal{H}_{\rm so} + \mathcal{H}_{\rm Coulomb}$ as the main Hamiltonian and hopping processes described by~$\mathcal{H}_{\rm tr}$ as a perturbation~\cite{book:Tyablikov1967MethodsIT}, 
\begin{equation}\label{eq:general_perturbation_theory}
    \mathcal{H}_{\rm eff} = \mathcal{P}\mathcal{H}_0\mathcal{P} - \mathcal{P}\mathcal{H}_{\rm tr}\mathcal{H}_0^{-1}(1 - \mathcal{P})\mathcal{H}_{\rm tr}\mathcal{P},
\end{equation}
where $\mathcal{P}$ is projection operator on the subspace with one electron at each lattice site. 
Obviously $\mathcal{P}\mathcal{H}_{\rm Coulomb}\mathcal{P} = 0$ and we %diagonalize the Coulomb interaction Hamiltonian $\mathcal{H}_{\rm Coulomb} $~(see~Eq.~\eqref{eq:main_H_Coulomb}) and 
neglect the impact of $\mathcal{H}_{\rm CF}$~[see Eq.~\eqref{eq:H_CF}] and $\mathcal{H}_{\rm so}$~[see Eq.~\eqref{eq:H_so}] on the eigenvalues and eigenfunctions of virtual (excited) states:
$$
\mathcal{P}\mathcal{H}_{\rm tr}\mathcal{H}_0^{-1}(1 - \mathcal{P})\mathcal{H}_{\rm tr}\mathcal{P} \approx
\mathcal{P}\mathcal{H}_{\rm tr}\mathcal{H}_{\rm Coulomb}^{-1}(1 - \mathcal{P})\mathcal{H}_{\rm tr}\mathcal{P}.
$$
This approximation is justified by that $|\Delta_{\rm CF}|, |\lambda| \ll U - 3J_{\rm H}$.

Our derivation generalizes the derivation of \cite{1978:Castellani} to the case of an arbitrary number of orbitals $N_{\rm d}$ (in the main text $N_{\rm d} = 3$). 
There have been also other approaches proposed to treat the same problem within the method of~irreducible operators~\cite{irkhin1993}.

There are two types of two-electron states at one site: {\it doubles}, characterized by double occupancy of a particular orbital (there are $N_{\rm d}$ such states), and  {\it pair-orbital} states [$2N_{\rm d}(N_{\rm d} - 1) = 4\times N_{\rm d}(N_{\rm d} - 1)/2$; factor 4 originates from the spin degeneracy].
These two sets of states  form invariant subspaces of $\mathcal{H}_{\rm Coulomb}$.

One can introduce basis functions for the subspace of doubles in the following way,
\begin{eqnarray}
\label{eq:E_dSigma_space_def}
E_{{\rm d}\Sigma} = U + (N_{\rm d} - 1)J_{\rm H}&:& A_{{\rm d}\Sigma}^\dag = \frac1{\sqrt{N_{\rm d}}}\sum_m c^\dag_{m\uparrow}c^\dag_{m\downarrow},\\
\label{eq:E_d_space_def}
E_{\mathrm{d}} = U - J_{\rm H}&:& A_{\mathrm{d}k}^\dag = \sum_{m}a^{(k)}_{m}c^\dag_{m\uparrow}c^\dag_{m\downarrow},
\end{eqnarray}
where $E_{\alpha}$ are the energies (eigenvalues of $\mathcal{H}_{\rm Coulomb}$) of the corresponding states, $k = 1, \ldots, N_{\rm d} - 1$, and the coefficients $a^{(k)}_{m}$ satisfy the relations $\sum_{m}a^{(k)}_{m} = 0$ and $\sum_{m}a^{(k)*}_{m}a^{(k')}_{m} = \delta_{kk'}$. We also use the notation $a^{(N_{\rm d})}_{m} = 1/\sqrt{N_{\rm d}}$, which corresponds to the wave function $A_{{\rm d}\Sigma}^\dag$, so that the matrix $a^{(k)}_{m}$ appears to be unitary.

For the subspace of pair orbitals, we have
\begin{equation}
A^\dag_{a:mm'} = \frac 1{\sqrt 2}\sum_{\sigma\sigma'}\sigma^o_{\sigma\sigma'}c^\dag_{m\sigma}c^\dag_{m'\sigma'}, \; \; \; m<m',
\end{equation}
with $ o = 0,x,y,z$, and there are once again two subspaces --- singlet~(S) and triplet~(T) ones with the energies
\begin{eqnarray}
\label{eq:general_multiplet.S}
E_{\mathrm{S}} = U'+J_{\rm H}&:&  o = y,\\
\label{eq:general_multiplet.T}
E_{\mathrm{T}}= U'-J_{\rm H}&:& o = 0,x,z.
\end{eqnarray}

As a whole, we have four eigenspaces, for which the projectors onto the corresponding excited states can be readily obtained,
\begin{eqnarray}
\label{eq:PdSigma_def}
\mathcal{P}_{{\rm d}\Sigma} &=& A^\dag_{{\rm d}\Sigma}A^{}_{{\rm d}\Sigma}, \\
\label{eq:Pd_def}
\tilde{\mathcal{P}}_{\mathrm{d}} &=& \sum_{k = 1}^{N_{\rm d}-1}A^\dag_{{\rm d}k}A^{}_{{\rm d}k}, \\
\label{eq:PS_def}
\mathcal{P}_{\mathrm{S}} &=& \sum_{m<m'}A^\dag_{y:mm'}A^{}_{y:mm'}, \\
\label{eq:PT_def}
\mathcal{P}_{\mathrm{T}} &=& \sum_{o=0,x,z}\sum_{m<m'}A^\dag_{a:mm'}A^{}_{a:mm'}.
\end{eqnarray}

In order to formulate the effective model in the second-order perturbation theory, we have to take the sum over all subspaces of excited states (given by $\alpha = {\rm d}\Sigma, \rm d, S, T$),
\begin{multline}\label{eq:H_kin_general}
\mathcal{H}_{\rm eff} = \mathcal{H}_{\rm CF} + \mathcal{H}_{\rm so} -\sum_\alpha E^{-1}_{\alpha}
\\
\sum_{ijm_1m_1'm_2m_2'\sigma\sigma'}t^{m_1m_2'}_{ij}
t^{m_1'm_2}_{ji}c^\dag_{im_1\sigma}c^{}_{im_2\sigma'}c^{}_{jm_2'\sigma}
\mathcal{P}_{j\alpha}c^\dag_{jm_1'\sigma'},
\end{multline}
where $\mathcal{P}_{j\alpha}$ is projector $\mathcal{P}_{\alpha}$ at site $j$. The last expression describes processes of the electron transfer from site $i$, orbital $m_2$, spin projection $\sigma$ to site $j$, orbital $m_1'$, and the same spin. Then, we project this excited state onto different subspaces and move the electron back to the initial site. It is convenient for further consideration, however, to rearrange $c$ operators in Eq.~\eqref{eq:H_kin_general} according to the site index. Here and below we assume that $\mathcal{H}_{\rm eff}$ acts on the subspace with one electron at each lattice site. 

Before proceeding to  calculating $c^{}_{m\sigma}\mathcal{P}_{\alpha}c^\dag_{m'\sigma'}$,  we present explicit expressions for some of the projectors via $c$ operators. From Eq.~(\ref{eq:PdSigma_def}), one can obtain
\begin{equation}\label{eq:PdSigma_expr}
\mathcal{P}_{{\rm d}\Sigma} = \frac1{N_{\rm d}}\sum_{mm'}c^\dag_{m\uparrow}c^\dag_{m\downarrow}c_{m'\downarrow}c_{m'\uparrow},
\end{equation}
while Eqs.~(\ref{eq:E_dSigma_space_def}) and (\ref{eq:PdSigma_def}) give
\begin{multline}
\tilde{\mathcal{P}}_{\mathrm{d}} = \sum_{mm'}\sum_{k = 1}^{N_{\rm d}-1}a^{(k)}_{m}a^{(k)*}_{m'}c^\dag_{m\uparrow}c^\dag_{m\downarrow}
c^{}_{m'\downarrow}c^{}_{m'\uparrow} \\
= \sum_{mm'}\left(\delta_{mm'} - 1/N_{\rm d}\right)c^\dag_{m\uparrow}c^\dag_{m\downarrow}c^{}_{m'\downarrow}c^{}_{m'\uparrow} = \mathcal{P}_{\mathrm{d}} - \mathcal{P}_{{\rm d}\Sigma},
\end{multline}
where
\begin{equation}\label{eq:P_d_def}
\mathcal{P}_{\mathrm{d}} = \sum_{m}c^\dag_{m\uparrow}c^{}_{m\uparrow}c^\dag_{m\downarrow}c^{}_{m\downarrow}
\end{equation}
is the projector onto subspace of doubles (d).  Correspondingly
for the triplet pair-orbital state [see definitions \eqref{eq:PS_def} and 
 \eqref{eq:PT_def}], we have
\begin{equation}\label{eq:PT}
\mathcal{P}_{\mathrm{T}} = \mathcal{B} + \mathcal{P}_{\mathrm{S}},
\end{equation}
and
\begin{equation}\label{eq:B_def}
\mathcal{B} = \sum_{m<m',\sigma\sigma'}c^\dag_{m\sigma}c^{}_{m\sigma'}
c^\dag_{m'\sigma'}c^{}_{m'\sigma},
\end{equation}
where the Fierz identity
\begin{equation}
\sum_{o = 0}^3\sigma^o_{\sigma_1\sigma_1'}\sigma^o_{\sigma_2\sigma_2'} = 2\delta_{\sigma_1\sigma_2'}\delta_{\sigma_2\sigma_1'}.
\end{equation}
was used.

Next, we calculate $c^{}_{m\sigma}\mathcal{P}_{\alpha}c^\dag_{m'\sigma'}$ from Eq.~\eqref{eq:H_kin_general} separately for each subspace (omitting the site index $j$). Eq.~(\ref{eq:PT}) allows us to consider $\mathcal{P}_{\alpha} = \mathrm{d}\Sigma, \mathrm{d}, \mathrm{S}$, and $\mathcal{B}$ (instead of T)
\begin{itemize}
\item $\alpha = \mathrm{d}\Sigma$. From Eq.~(\ref{eq:PdSigma_expr}), we readily find
\begin{equation}
c^{}_{m\sigma}\mathcal{P}_{\mathrm{d}\Sigma}c^\dag_{m'\sigma'} = \frac1{N_{\rm d}}\sum_{m_1m_1'}\left[c^{}_{m\sigma},c^\dag_{m_1\uparrow}
c^\dag_{m_1\downarrow}\right]\left[c_{m_1'\downarrow}c_{m_1'\uparrow},
c^\dag_{m'\sigma'}\right].
\end{equation}

Here and below, the terms, which are not bilinear form in the Fermi operators, are omitted since they are eventually projected out of the considered state. Thus, we obtain
\begin{equation}\label{eq:dSigma_projection}
c^{}_{m\sigma}\mathcal{P}_{\mathrm{d}\Sigma}c^\dag_{m'\sigma'} = \left(1/{N_{\rm d}}\right)\gamma_\sigma\gamma_{\sigma'} c^\dag_{m\bar\sigma}c^{}_{m'\bar\sigma'},
\end{equation}
where $\gamma_\uparrow = +1, \gamma_\downarrow = -1$, and finally one gets the expression entering Eq.~(\ref{eq:H_kin_general}) for ${\rm d}\Sigma$ subspace
\begin{multline}
\sum_{\sigma\sigma'}c^\dag_{im_1\sigma}c^{}_{im_2\sigma'}
c^{}_{jm_2'\sigma}\mathcal{P}_{j{\rm d}\Sigma}c^\dag_{jm_1'\sigma'}
\\
= \left(1/{N_{\rm d}}\right)\sum_{\sigma\sigma'}c^\dag_{im_1\sigma}
c^{}_{im_2\sigma'}\gamma_\sigma\gamma_{\sigma'} c^\dag_{jm_2'\bar\sigma}c^{}_{jm_1'\bar\sigma'} \\
= \left(2/{N_{\rm d}}\right)\left(\mathcal{S}^{(0)}_{im_1m_2}\mathcal{S}^{(0)}_{jm_2'm_1'} - \bm{\mathcal{S}}^{}_{im_1m_2}\bm{\mathcal{S}}^{}_{jm_2'm_1'}\right),
\end{multline}
where $\mathcal{S}^{(o)}_{imm'}$ is defined by~Eq.~\eqref{eq:S_mm'_def}. 
%we used Eq.~(\ref{eq:spin_combination_gamma}) to rewrite the final expression via a generalized spin operator
Without the orbital index, all its components coincide with the conventional spin (and number of particles) operators. The conventional charge and spin operators are obtained by taking trace of $\mathcal{S}^{(o)}_{imm'}$ over orbital indices.

\item $\alpha = \mathrm{d}$.
Following the same strategy, one can find using Eq.~(\ref{eq:P_d_def}) 
\begin{multline}
c^{}_{m\sigma}\mathcal{P}_{\mathrm{d}}c^\dag_{m'\sigma'} = \sum_{m_1}\left[c^{}_{m\sigma},c^\dag_{m_1\uparrow}
c^\dag_{m_1\downarrow}\right]\left[c^{}_{m_1\downarrow}
c^{}_{m_1\uparrow},c^\dag_{m'\sigma'}\right]
\\
= \delta_{mm'}\gamma_\sigma\gamma_{\sigma'} c^\dag_{m\bar\sigma}c^{}_{m\bar\sigma'}.
\end{multline}
and finally, the expression entering ~Eq.~\eqref{eq:H_kin_general} in terms of generalized spin operators transforms to
\begin{multline}
\sum_{\sigma\sigma'}c^\dag_{im_1\sigma}c^{}_{im_2\sigma'}
c^{}_{jm_2'\sigma}\tilde{\mathcal{P}}_{j\mathrm{d}}c^\dag_{jm_1'\sigma'}
\\
= \left(\delta_{m_1'm_2'}-1/N_{\rm d}\right)\sum_{\sigma\sigma'}\gamma_\sigma\gamma_{\sigma'}
c^\dag_{im_1\sigma}c^{}_{im_2\sigma'}c^\dag_{jm_2'\bar\sigma}
c^{}_{jm_1'\bar\sigma'}\\
=2\left(\delta_{m_1'm_2'}-1/N_{\rm d}\right)\left(\mathcal{S}^{(0)}_{im_1m_2}\mathcal{S}^{(0)}_{jm_2'm_1'} - \bm{\mathcal{S}}^{}_{im_1m_2}\bm{\mathcal{S}}^{}_{jm_2'm_1'}\right).
\end{multline}
\item $\alpha = \mathcal{B}$.
\begin{multline}
c^{}_{m\sigma}\mathcal{B}c^\dag_{m'\sigma'} \\
= \sum_{m_1<m_1',\sigma_1\sigma_1'}
\left[c^{}_{m\sigma},c^\dag_{m_1\sigma_1}c^\dag_{m_1'\sigma_1'}\right]
\left[c^{}_{m_1'\sigma_1}c^{}_{m_1\sigma_1'},c^\dag_{m'\sigma'}\right] \\
= \sum_{m_1<m_1',\sigma_1\sigma_1'}
\left(\delta_{mm_1}\delta_{\sigma\sigma_1}c^\dag_{m_1'\sigma_1'} - \delta_{mm_1'}\delta_{\sigma\sigma_1'}c^\dag_{m_1\sigma_1}\right)
\\
\left(\delta_{m'm_1}\delta_{\sigma'\sigma_1'}c_{m_1'\sigma_1} - \delta_{m'm_1'}\delta_{\sigma'\sigma_1}c_{m_1\sigma_1'}\right).
\end{multline}
Using symmetry of this expression with respect to the orbital index exchange $m_1\leftrightarrow m_1'$ we obtain
\begin{multline}
c^{}_{m\sigma}\mathcal{B}c^\dag_{m'\sigma'} = \delta_{mm'}\sum_{m_1\ne m}c^{\dag}_{m_1\sigma'}c^{}_{m_1\sigma} 
\\
- \delta_{\sigma\sigma'}(1 - \delta_{mm'})\sum_{\sigma_1}c^{\dag}_{m'\sigma_1}c^{}_{m\sigma_1}.
\end{multline}
%\begin{multline}
%\sum_{\sigma\sigma'}c^\dag_{im_1\sigma}c^{}_{im_2\sigma'}
%c^{}_{jm_2'\sigma}\mathcal{B}_{j}c^\dag_{jm_1'\sigma'}= \sum_{\sigma\sigma'}c^\dag_{im_1\sigma}c^{}_{im_2\sigma'}
%\\
%\left[\delta_{m_1'm_2'}\sum_{m\ne m_1'}c^{\dag}_{jm\sigma'}c^{}_{jm\sigma} - \delta_{\sigma\sigma'}(1 - \delta_{m_1'm_2'})\sum_{\sigma_1}c^{\dag}_{jm_1'\sigma_1}c^{}_{jm_2'\sigma_1}\right].
%\end{multline}

Finally  we obtain
\begin{multline}
\sum_{\sigma\sigma'}c^\dag_{im_1\sigma}c^{}_{im_2\sigma'}
c^{}_{jm_2'\sigma}\mathcal{B}_{j}c^\dag_{jm_1'\sigma'} \\
= 2\delta_{m_1'm_2'}\sum_{m\ne m_1'}\left(\mathcal{S}^{(0)}_{i;m_1m_2}\mathcal{S}^{(0)}_{j;mm} + \bm{\mathcal{S}}_{i;m_1m_2}\bm{\mathcal{S}}_{j;mm}\right)
\\
- 4(1 - \delta_{m_1'm_2'})\mathcal{S}^{(0)}_{i;m_1m_2}\mathcal{S}^{(0)}_{j;m_1'm_2'}.
\end{multline}

\item $\alpha = \mathrm{S}$.
\begin{equation}
c^{}_{m\sigma}\mathcal{P}_{\mathrm{S}}c^\dag_{m'\sigma'} = \sum_{m_1<m_1'}\left[c^{}_{m\sigma},A^\dag_{y:m_1m_1'}\right]
\left[A^{}_{y:m_1m_1'},c^\dag_{m'\sigma'}\right]
\end{equation}
and since
\begin{multline}
\left[A^{}_{y:m_1m_1'},c^\dag_{m\sigma}\right] = (\I/\sqrt2)\left[c_{m'_1\downarrow}c_{m_1\uparrow} - c_{m'_1\uparrow}c_{m_1\downarrow},c^\dag_{m\sigma}\right]
\\
= \I(\gamma_\sigma/\sqrt2)\left(
\delta_{mm_1}c_{m'_1\bar\sigma} + \delta_{mm_1'}c_{m_1\bar\sigma}
\right),
\end{multline}
we find using $m_1\leftrightarrow m_1'$ symmetry that
\begin{multline}
c^{}_{m\sigma}\mathcal{P}_{\mathrm{S}}c^\dag_{m'\sigma'} =
\frac14\gamma_\sigma\gamma_{\sigma'}
\sum_{m_1\ne m_1'}
\left(
\delta_{mm_1}c^\dag_{m'_1\bar\sigma} + \delta_{mm_1'}c^\dag_{m_1\bar\sigma}
\right)
\\
\times
\left(
\delta_{m'm_1}c_{m'_1\bar\sigma'} + \delta_{m'm_1'}c_{m_1\bar\sigma'}
\right).
\end{multline}
%This expression can be rewritten as
%\begin{multline}
%c^{}_{m\sigma}\mathcal{P}_{\mathrm{S}}c^\dag_{m'\sigma'} = \frac14\gamma_\sigma\gamma_{\sigma'}
%\\
%\sum_{m_1, m_1'}
%\left(
%\delta_{mm_1}c^\dag_{m'_1\bar\sigma} + \delta_{mm_1'}c^\dag_{m_1\bar\sigma}
%\right)
%\left(
%\delta_{m'm_1}c_{m'_1\bar\sigma'} + \delta_{m'm_1'}c_{m_1\bar\sigma'}
%\right)
%\\
%- \gamma_\sigma\gamma_{\sigma'}\delta_{mm'}c^\dag_{m\bar\sigma}c^{}_{m\bar\sigma'}
%\label{eq:S_projection}
%= \frac12\gamma_\sigma\gamma_{\sigma'} \times
%\\
%\left(
%\delta_{mm'}\sum_{m_1}\left(1-2\delta_{mm_1}\right)c^\dag_{m_1\bar\sigma}
%c^{}_{m_1\bar\sigma'}
%+ c^\dag_{m'\bar\sigma}c^{}_{m\bar\sigma'}
%\right).
%\end{multline}

Finally, the expression entering Eq.~(\ref{eq:H_kin_general}) for the subspace S is given by
\begin{multline}
\sum_{\sigma\sigma'}c^\dag_{im_1\sigma}c^{}_{im_2\sigma'}
c^{}_{jm_2'\sigma}\mathcal{P}_{j\mathrm{S}}c^\dag_{jm_1'\sigma'} \\
= \delta_{m_1'm_2'}\sum_m(1-2\delta_{m_1'm})
\left(\mathcal{S}^{(0)}_{im_1m_2}\mathcal{S}^{(0)}_{jmm} - \bm{\mathcal{S}}^{}_{im_1m_2}\bm{\mathcal{S}}^{}_{jmm}\right)
\\
+ \mathcal{S}^{(0)}_{im_1m_2}\mathcal{S}^{(0)}_{jm_1'm_2'} - \bm{\mathcal{S}}^{}_{im_1m_2}\bm{\mathcal{S}}^{}_{jm_1'm_2'}.
\end{multline}
\end{itemize}
Using Eq.~\eqref{eq:PT}, we get analogously for the T subspace
%\begin{multline}
%\sum_{\sigma\sigma'}c^\dag_{im_1\sigma}c^{}_{im_2\sigma'}c^{}_{jm_2'\sigma}\mathcal{P}_{j\mathrm{T}}c^\dag_{jm_1'\sigma'} =
%\\
%2\delta_{m_1'm_2'}\sum_{m\ne m_1'}\left(\mathcal{S}^{(0)}_{i;m_1m_2}\mathcal{S}^{(0)}_{j;mm} + \bm{\mathcal{S}}_{i;m_1m_2}\bm{\mathcal{S}}_{j;mm}\right) - 4(1 - \delta_{m_1'm_2'})\mathcal{S}^{(0)}_{i;m_1m_2}\mathcal{S}^{(0)}_{j;m_1'm_2'}
%\\
%+
%\delta_{m_1'm_2'}\sum_{m\ne m_1'}\left(\mathcal{S}^{(0)}_{im_1m_2}\mathcal{S}^{(0)}_{jmm} - \bm{\mathcal{S}}^{}_{im_1m_2}\bm{\mathcal{S}}^{}_{jmm}\right)
%+ \mathcal{S}^{(0)}_{im_1m_2}\mathcal{S}^{(0)}_{jm_1'm_2'} - \bm{\mathcal{S}}^{}_{im_1m_2}\bm{\mathcal{S}}^{}_{jm_1'm_2'}
%\\
%-
%\delta_{m_1'm_2'}\left(\mathcal{S}^{(0)}_{im_1m_2}\mathcal{S}^{(0)}_{jm_1'm_1'} - \bm{\mathcal{S}}^{}_{im_1m_2}\bm{\mathcal{S}}^{}_{jm_1'm_1'}\right)
%\end{multline}
\begin{multline}
\sum_{\sigma\sigma'}c^\dag_{im_1\sigma}c^{}_{im_2\sigma'}
c^{}_{jm_2'\sigma}\mathcal{P}_{j\mathrm{T}}c^\dag_{jm_1'\sigma'} =
\\
\delta_{m_1'm_2'}\sum_{m}\left(3\mathcal{S}^{(0)}_{i;m_1m_2}\mathcal{S}^{(0)}_{j;mm} + \bm{\mathcal{S}}_{i;m_1m_2}\bm{\mathcal{S}}_{j;mm}\right) \\
- 3\mathcal{S}^{(0)}_{i;m_1m_2}\mathcal{S}^{(0)}_{j;m_1'm_2'} - \bm{\mathcal{S}}_{i;m_1m_2}\bm{\mathcal{S}}_{j;m_1'm_2'}.
\end{multline}

Combining all the results together and summing over the excited states, we arrive at a final expression for the effective Hamiltonian in terms of the  generalized spin operators,
\begin{multline}\label{eq:final_exchange}
\mathcal{H}_{\rm eff} = \mathcal{H}_{\rm CF} + \mathcal{H}_{\rm so}
-\sum_{ijm_1m_1'm_2m_2'}t^{m_1m_2'}_{ij}t^{m_1'm_2}_{ji}
\\
\left(-\frac{2J_{\rm d}}{(U + (N_{\rm d}-1)J_{\rm d})(U - J_{\rm d})}
\right.
\\
\left.
\left(\mathcal{S}^{(0)}_{im_1m_2}\mathcal{S}^{(0)}_{jm_2'm_1'} - \bm{\mathcal{S}}^{}_{im_1m_2}\bm{\mathcal{S}}^{}_{jm_2'm_1'}\right)
\right.
\\
\left.
+
2\delta_{m_1'm_2'}\left(\frac1{U - J_{\rm d}}-\frac1{U' + J_{\rm H}}\right)
\right.
\\
\left.
\times
\left(\mathcal{S}^{(0)}_{im_1m_2}\mathcal{S}^{(0)}_{jm_2'm_1'} - \bm{\mathcal{S}}^{}_{im_1m_2}\bm{\mathcal{S}}^{}_{jm_2'm_1'}\right)
\right.
\\
\left.
+\delta_{m_1'm_2'}
\left(
\left(
\frac1{U' + J_{\rm H}} + \frac3{U' - J_{\rm H}}\right)\mathcal{S}^{(0)}_{im_1m_2}\sum_m\mathcal{S}^{(0)}_{jmm}
\right.\right.
\\
\left.\left.
+ \left(\frac1{U' - J_{\rm H}} - \frac1{U' + J_{\rm H}}\right)\bm{\mathcal{S}}^{}_{im_1m_2}\sum_m\bm{\mathcal{S}}^{}_{jmm}
\right)
\right.
\\
\left.
- \left(\frac3{U' - J_{\rm H}} - \frac1{U' + J_{\rm H}}\right)\mathcal{S}^{(0)}_{im_1m_2}\mathcal{S}^{(0)}_{jm_1'm_2'} -
\right.
\\
\left.
\left(\frac1{U' - J_{\rm H}} + \frac1{U' + J_{\rm H}}\right)\bm{\mathcal{S}}^{}_{im_1m_2}\bm{\mathcal{S}}^{}_{jm_1'm_2'}
\right).
\end{multline}

For the hopping parameters, which are diagonal in the orbital space $t^{mm'}_{ij} = \delta_{mm'}t^{m}_{ij}$, using the Kanamori parametrization~\cite{Kanamori1963} $U' = U - 2J_{\rm H}$, $J_{\rm d} = J_{\rm H}$ and taking into account that $\sum_m\mathcal{S}^{(0)}_{jmm} = 1/2$, one obtains
\begin{multline}
\label{eq:kinetic_exchange_H_final}
\mathcal{H}_{\rm eff} = \mathcal{H}_{\rm CF} + \mathcal{H}_{\rm so}
\\
-\sum_{ijm}t^{m}_{ij}t^{m}_{ji}
\left[
\frac12\left( \frac1{U - J_{\rm H}} + \frac3{U - 3J_{\rm H}}\right)
\mathcal{S}^{(0)}_{imm}
\right.
\\
\left.
+ \left(\frac1{U - 3J_{\rm H}} 
- \frac1{U - J_{\rm H}} \right) \bm{\mathcal{S}}^{}_{imm}\mathbf{S}_j \right] 
\\
+
\sum_{ijm_1m_2}t^{m_1}_{ij}t^{m_2}_{ji}
 \left[ \frac{2J_{\rm d}}{(U + (N_{\rm d}-1)J_{\rm H})(U - J_{\rm H})} \right.
\\
\left.
\times
\left(\mathcal{S}^{(0)}_{im_1m_2}\mathcal{S}^{(0)}_{jm_1m_2} - \bm{\mathcal{S}}^{}_{im_1m_2}\bm{\mathcal{S}}^{}_{jm_1m_2}\right) 
\right.
\\
+
\left.
\left(\frac3{U - 3J_{\rm H}} - \frac1{U - J_{\rm H}}\right)\mathcal{S}^{(0)}_{im_1m_2}\mathcal{S}^{(0)}_{jm_2m_1}
\right.
\\
\left.
+ \left(\frac1{U - 3J_{\rm H}} + \frac1{U - J_{\rm H}}\right)\bm{\mathcal{S}}^{}_{im_1m_2}\bm{\mathcal{S}}^{}_{jm_2m_1}
\right],
\end{multline}
where $\mathbf{S}_i$ is defined by Eq.~\eqref{S-def}.
%= \sum_m\bm{\mathcal{S}}^{}_{imm}$ is a conventional (total) site spin operator.

\begin{figure}[t]
   \centering  \includegraphics[angle = -90, width=0.5\textwidth]{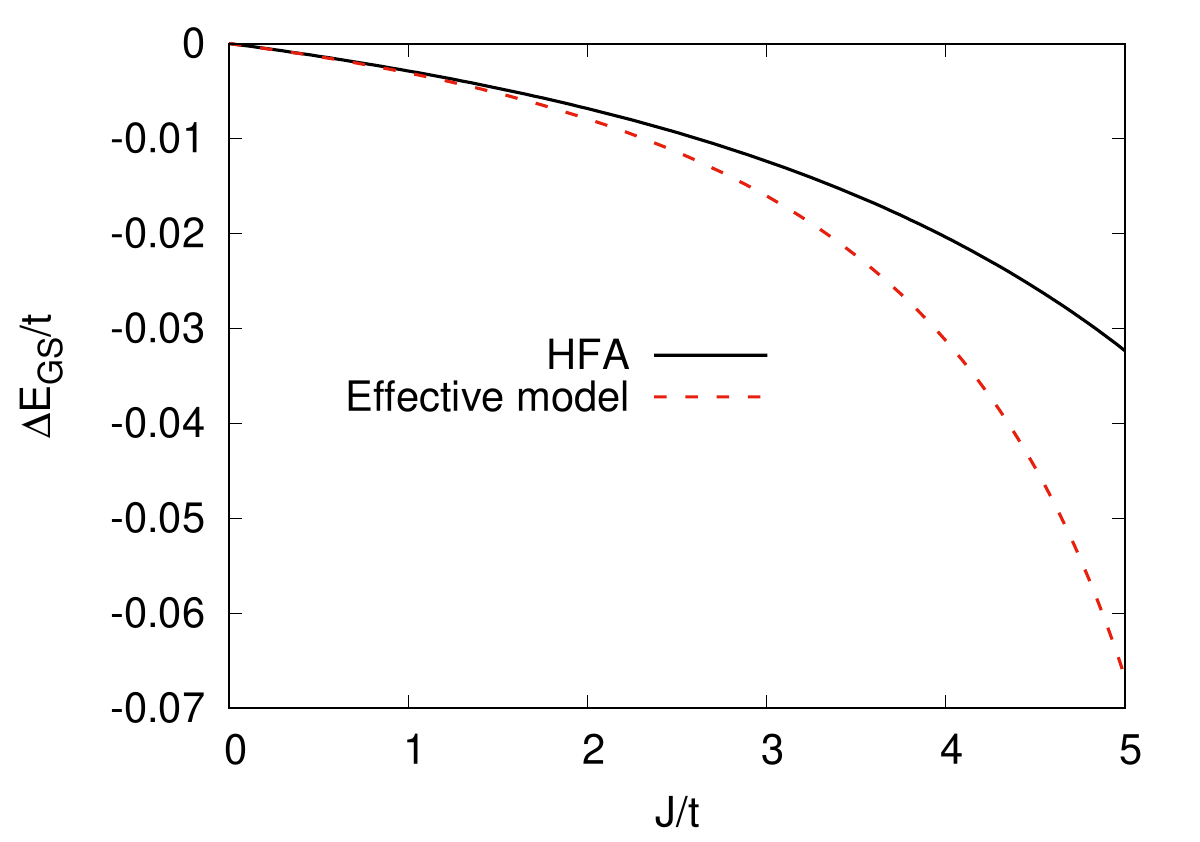}  
   \caption{ 
   \label{Fig:AFMoe-TAFM} 
   Total energy difference between $\texttt{AFM-TM}$ and $\texttt{AFO-eO}$ states for $U=20t$, $\Delta_{\rm CF} = 0$, $\lambda = 0$ as obtained by the direct calculation using Hartree-Fock methods (black solid line) and effective model for kinetic exchange within the mean-field approximation  developed in Appendix~A (red dashed line); see~Eq.~\eqref{eq:H_kin_non-sym_exchange_term}. 
   }
\end{figure}

Below we consider in detail an application of~Eq.~\eqref{eq:kinetic_exchange_H_final} for the calculation of the difference between energies of the \texttt{AFM-TM} and \texttt{AFO-eO} states (see Fig.~\ref{Fig:AFMoe-TAFM}). 

Using the mean-field approximation for the Hamiltonian \eqref{eq:kinetic_exchange_H_final} is equivalent to the assumption that $\Phi$ has a two-sublattice form,
\begin{equation}
\Phi = \prod_{i_1\in \rm A}\prod_{i_2\in \rm B} a^\dag_{i_1}b^\dag_{i_2}|0>,
\end{equation}
where $|0>$ is vacuum, and site one-electron state $a$ corresponds to sublattice A and $b$  to sublattice B. Then $E_{\rm kin} = \langle\Phi|\mathcal{H}_{\rm kin}|\Phi\rangle$. Under this approximation we have to replace charge-spin operators in Eq.~\eqref{eq:H_kin_diag} by their averages. So we have to calculate $\langle\mathcal{S}^{(0)}_{imm'}\rangle$ and $\langle\bm{\mathcal{S}}^{}_{imm'}\rangle$ to calculate the energy of the state $\Phi$.  

Omitting site index $i$ and writing generally for $d = a, b$
\begin{equation}
d^\dag = \sum_{m\sigma}u^d_{m\sigma}c^\dag_{m\sigma},   
\end{equation}
with $u_{m\sigma}$ being arbitrary normalized coefficients, we get
\begin{equation}\label{eq:Sa_value}
\langle\mathcal{S}^{(o)}_{mm'}\rangle_d = \frac12\sum_{\sigma\sigma'}\sigma^{(o)}_{\sigma\sigma'}(u^d_{m\sigma})^*u^d_{m'\sigma'}.
\end{equation}

%Below we omit angular brackets for brevity.
Since the $xy$ orbital is not occupied we simplify the expression (\ref{eq:kinetic_exchange_H_final}) for energy taking into account in the sum over $m_1,m_2$ only diagonal terms ($m_1 = m_2$),
\begin{multline}\label{eq:H_kin_diag}
E_{\rm kin}(\Phi) =
\sum_{ij,m_1\ne xy}|t^{m_1}_{ij}|^2
\\
\left(-\frac{\langle s^{(0)}_{im_1}\rangle}2\left(
\frac1{U - J_{\rm H}} + \frac3{U - 3J_{\rm H}}\right) 
\right.
\\
\left.
- \left(\frac1{U - 3J_{\rm H}} - \frac1{U - J_{\rm H}}\right)\langle\mathbf{s}_{im_1}\rangle\langle\mathbf{S}^{}_{j}\rangle
\right.
\\
+
\left.
\frac{2J_{\rm H}}{(U + 2J_{\rm H})(U - J_{\rm H})}
\left(\langle s^{(0)}_{im_1}\rangle\langle s^{(0)}_{jm_1}\rangle - \langle\mathbf{s}_{im_1}\rangle\langle\mathbf{s}_{jm_1}\rangle\right)
\right.
\\
\left.
+
\left(\frac3{U - 3J_{\rm H}} - \frac1{U - J_{\rm H}}\right)\langle s^{(0)}_{im_1}\rangle \langle s^{(0)}_{jm_1}\rangle 
\right.
\\
\left.
+ \left(\frac1{U - 3J_{\rm H}} 
+ \frac1{U - J_{\rm H}}\right)\langle\mathbf{s}_{im_1}\rangle\langle\mathbf{s}_{jm_1}\rangle
\right),
\end{multline}
where the spin $s^{(o)}_{im}\equiv \mathcal{S}^{(o)}_{imm}$ operator of an orbital $m$ at a site $i$ is introduced.  

Now we consider separately $E_{\rm kin}(\Phi)$ for \texttt{TAFM} and \texttt{oe-AFM} states. 
\begin{itemize}
\item \texttt{AFM-TM} order. 
We assume that 
\begin{eqnarray}
a^\dag &=& -\frac1{\sqrt2}\left(\I c^\dag_{xz\uparrow} + c^\dag_{yz\uparrow}\right), \\ 
b^\dag &=& -\frac1{\sqrt2}\left(\I c^\dag_{xz\downarrow} - c^\dag_{yz\downarrow}\right),
\end{eqnarray}
so $u^{a,b}_{m\sigma} = -\I\delta_{m,xz}/\sqrt2 \mp \delta_{m,xy}/\sqrt2$.

%For $i\in \rm A$ 
%$$
%\langle\mathcal{S}^{(o)}_{imm'}\rangle = \frac14\sigma^{(o)}_{\uparrow\uparrow}(-\I\delta_{m,xz} + \delta_{m,yz})(\I\delta_{m',xz} + \delta_{m',yz}).
%$$
%Analogously, for $i\in \rm B$  we get 
%$$
%\langle\mathcal{S}^{(o)}_{imm'}\rangle = \frac14\sigma^{(o)}_{\downarrow\downarrow}(-\I\delta_{m,xz} - \delta_{m,yz})(\I\delta_{m',xz} - \delta_{m',yz}).
%$$
We get from Eq.~\eqref{eq:Sa_value}
\begin{eqnarray}
\langle s^{(o)}_m\rangle_{\rm A} &=& \frac14\sigma^{(o)}_{\uparrow\uparrow}(1-\delta_{m,xy}), \\
\langle s^{(o)}_m\rangle_{\rm B} &=& \frac14\sigma^{(o)}_{\downarrow\downarrow}(1-\delta_{m,xy}).
\end{eqnarray}
\item \texttt{AFO-eO} order.
\begin{eqnarray}
a^\dag &=& -\frac12\left(\I c^\dag_{xz\uparrow} + c^\dag_{yz\uparrow} + \I c^\dag_{xz\downarrow} - c^\dag_{yz\downarrow}\right), \\ 
b^\dag &=& -\frac12\left(\I c^\dag_{xz\uparrow} + c^\dag_{yz\uparrow} - \I c^\dag_{xz\downarrow} + c^\dag_{yz\downarrow}\right),
\end{eqnarray}
so $u^a_{m\sigma} = -\I\delta_{m,xz}/2 - \gamma_\sigma\delta_{m,yz}/2$, $u^b_{m\sigma} = -\I \gamma_\sigma\delta_{m,xz}/2 - \delta_{m,yz}/2$, $\gamma_\sigma = +1(-1)$ for $\sigma = \uparrow(\downarrow)$.
%$$
%\langle\mathcal{S}^{(o)}_{imm'}\rangle = \frac18\sum_{\sigma\sigma'}\sigma^{(o)}_{\sigma\sigma'}(-\I\delta_{m,xz} + \gamma_\sigma\delta_{m,yz})(\I\delta_{m',xz} + \gamma_{\sigma'}\delta_{m',yz}).
%$$
%Analogously, for $i\in \rm B$  we assume that $\beta_{m\sigma} = \gamma_{m\sigma}$ and 
%$$
%\langle\mathcal{S}^{(o)}_{imm'}\rangle = \frac18\sum_{\sigma\sigma'}\sigma^{(o)}_{\sigma\sigma'}(-\I\gamma_\sigma\delta_{m,xz} + \delta_{m,yz})(\I\gamma_{\sigma'}\delta_{m',xz} + \delta_{m',yz}).
%$$
We get analogously
\begin{multline}
\langle s^{(o)}_m\rangle_{\rm A,B}  = \frac14\delta_{a,0}(\delta_{m,xz} + \delta_{m,yz})
\\
\pm \frac14\delta_{a,x}(\delta_{m,xz} - \delta_{m,yz}).
\end{multline}
\end{itemize}
Note that for neighboring sites $i$ and $j$, both orders $\langle s^{(0)}_{im}\rangle\langle s^{(0)}_{jm}\rangle = 1/16$, $\langle \mathbf{s}^{}_{im}\rangle\langle \mathbf{s}^{}_{jm}\rangle = -1/16$. Therefore all terms in Eq.~(\ref{eq:H_kin_diag}) coincide besides underlined term. 
In the case \texttt{AFM-TM} order we get $\langle \mathbf{S}^{}_{i}\rangle  = (-1)^i\mathbf{e}_z/2$ with unit vector directed along $z$ axis $\mathbf{e}_z$; therefore $\langle\mathbf{s}^{}_{im_1}\rangle\langle\mathbf{S}^{}_{j}\rangle = -1/8$. This term has a different nature than another terms in Eq.~(\ref{eq:H_kin_diag}) since here \textit{total} spin exhibits \textit{ferromagnetic} exchange interaction with spin of neighbor orbitals (indexed by $m_1$). In the case of \texttt{AFO-eO} order $\langle\mathbf{S}^{}_{j}\rangle  = 0$. We find that ferromagnetic exchange interaction increases the energy of \texttt{TAFM} order making it less favorable than \texttt{AFO-eO} order due to a delicate effect of ferromagnetic exchange interaction. 

%%%%%%%%%%%%%%%%%%%%%%%%%%%%%%%%%%
\section{Inter-orbital Coulomb interaction Hamiltonian within HFA}
In this appendix we present a HFA treatment of the local Coulomb Hamiltonian $\mathcal{H}_{\rm Coulomb}$~\eqref{eq:main_H_Coulomb}, Sec.~1, and  use this for the derivation of the total HFA Hamiltonian, see Eq.~\eqref{eq:H_MF_total},  Sec.~2. 

\subsection{Coulomb interaction Hamiltonian treatment}
We write the Coulomb interaction Hamiltonian omitting for brevity site index
\begin{multline}
\mathcal{H}_{\rm Coulomb} = \frac{U}2\sum_{m\sigma}c^\dag_{m\sigma}c^\dag_{m\bar\sigma}c^{}_{m\bar\sigma}c^{}_{m\sigma} \\
+\frac{J_{\rm d}}2\sum_{m\ne m';\sigma}c^\dag_{m\sigma}c^\dag_{m\bar\sigma}c^{}_{m'\bar\sigma}c^{}_{m'\sigma}  \\
+ \frac{U'}2\sum_{m\neq m'\sigma\sigma'}c^\dag_{m\sigma}c^\dag_{m'\sigma'}c^{}_{m'\sigma'}c^{}_{m\sigma}
\\
+ \frac{J_{\rm H}}2\sum_{m\ne m';\sigma\sigma'}c^\dag_{m\sigma}c^\dag_{m'\sigma'}c^{}_{m\sigma'}c^{}_{m'\sigma}.
\end{multline}
This sum is rewritten as
\begin{equation}
\mathcal{H}_{\rm Coulomb} =\sum_{\beta=U,J_{\rm d},U',J_{\rm H}} \mathcal{H}_{\rm int}\left[W^{(\beta)}\right],
\end{equation}
where 
\begin{multline}
    \mathcal{H}_{\rm int}[W] = 
\\
\frac12\sum_{mm'm_1m_1'\sigma\sigma'}
W_{\sigma\sigma'}(mm';m_1m_1')c^\dag_{m\sigma}c^\dag_{m'\sigma'}c^{}_{m_1'\sigma'}c^{}_{m_1\sigma}
\end{multline}
and
\begin{eqnarray}
\label{eq:W_U}
W^{(U)}_{\sigma\sigma'} &=& 
U\cdot\left(1-\delta_{\sigma\sigma'}\right)\delta_{mm'}\delta_{mm_1}\delta_{mm_1'}, \\
\label{eq:W_U'}
W^{(U')}_{\sigma\sigma'} &=& U'\left(1-\delta_{mm'}\right)\delta_{mm_1}\delta_{m'm_1'}, \\
\label{eq:W_J}
W^{(J_{\rm H})}_{\sigma\sigma'} &=& J_{\rm H}\left(1-\delta_{mm'}\right)\delta_{mm_1'}\delta_{m'm_1}, \\
\label{eq:W_Jd}
W^{(J_{\rm d})}_{\sigma\sigma'} &=& 
 J_{\rm d}\left(1-\delta_{\sigma\sigma'}\right)\cdot\left(1-\delta_{mm_1}\right)\delta_{mm'}\delta_{m_1m_1'},
\end{eqnarray}
where arguments of $W^{(\beta)}(mm';m_1m_1')$ are omitted for brevity. 

The HFA treatment of all terms is performed in an equal way ($\mathcal{H}_{\rm int}[W^{(\beta)}] \rightarrow \mathcal{H}^{\rm HFA}_{\rm int}[W^{(\beta)}]$) using the Wick theorem:
\begin{multline}\label{eq:H_int^HFA[W]}
\mathcal{H}^{\rm HFA}_{\rm int}\left[W^{(\beta)}\right]
= \sum_{mm'm_1m_1'\sigma\sigma'}W^{(\beta)}_{\sigma\sigma'}(mm';m_1m_1')
\\
\left(
c^\dag_{m\sigma}c^{}_{m_1\sigma}\langle c^\dag_{m'\sigma'}c^{}_{m_1'\sigma'}\rangle  - c^\dag_{m\sigma} c^{}_{m_1'\sigma'}\langle c^\dag_{m'\sigma'}c^{}_{m_1\sigma}\rangle\right) - E^{(\beta)}_{\rm DC},
\end{multline}
where $E^{(\beta)}_{\rm DC}$ is introduced to avoid double counting and equals half of the statistical average of the first two terms.

%Below we use spin identities 
%\begin{eqnarray}
%\label{eq:spin_combination_equal}
%\sum_\sigma c^\dag_{1\sigma}c^{}_{1\sigma}c^\dag_{2\sigma}c^{}_{2\sigma} &=& 2(s^0_1s^0_2 + s^z_1s^z_2),\\
%\label{eq:spin_combination_opposite}
%\sum_\sigma c^\dag_{1\sigma}c^{}_{1\sigma}c^\dag_{2\bar\sigma}c^{}_{2\bar\sigma} &=& 2(s^0_1s^0_2 - s^z_1s^z_2),\\
%\label{eq:spin_combination_opposite.2}
%\sum_\sigma c^\dag_{1\sigma}c^{}_{1\bar\sigma}c^\dag_{2\bar\sigma}c^{}_{2\sigma} &=& s^+_1s^-_2 + s^-_1s^+_2,\\
%\label{eq:spin_combination_double}
%\sum_{\sigma\sigma'} c^\dag_{1\sigma}c^{}_{1\sigma'}c^\dag_{2\sigma'}c^{}_{2\sigma} &=& 2(s^0_1s^0_2 + \mathbf{s}_1\mathbf{s}_2), \\
%\label{eq:spin_combination_gamma}
%\sum_{\sigma\sigma'} \gamma_\sigma\gamma_{\sigma'} c^\dag_{1\sigma}c^{}_{1\sigma'}c^\dag_{2\bar\sigma}c^{}_{2\bar\sigma'} &=& 2(s^0_1s^0_2 - \mathbf{s}_1\mathbf{s}_2),
%\end{eqnarray}
%where $s^{(o)}_{mm'} = \frac12\sum_{\sigma\sigma'}\sigma^{(o)}_{\sigma\sigma'}c^\dag_{m\sigma}c^{}_{m'\sigma'}$ are spin-like objects taken on a pair of arbitrary orbitals $m$ and $m'$. 

For the case $\beta = U, J_{\rm d}$ the spin projection dependence reads $W^{(\beta)}_{\sigma\sigma'}(mm';m_1m_1') \propto(1-\delta_{\sigma\sigma'})$ and we rewrite $\sigma$ sums in Eq.~\eqref{eq:H_int^HFA[W]} through generalized density and spin operators, see Eq.~\eqref{eq:S_mm'_def},
%we use Eq.~\eqref{eq:spin_combination_opposite} (Eq.~\eqref{eq:spin_combination_opposite.2}) for the first (second) term in~Eq.~\eqref{eq:H_int^HFA[W]}. We get
\begin{multline}\label{eq:H_int^HFA[W=U,Jd]}
\mathcal{H}^{\rm HFA}_{\rm int}\left[W^{(\beta)}\right]
= \sum_{mm'm_1m_1'}W^{(\beta)}_{\uparrow\downarrow}(mm';m_1m_1')
\\
\left(
2\mathcal{S}^{(0)}_{mm_1}\no_{m'm_1'} - 
2{\mathcal{S}}^z_{mm_1}\mathfrak{m}^z_{m'm_1'}  
\right.
\\
\left.
- \mathcal{S}^{+}_{mm_1'}\mathfrak{m}^-_{m'm_1} - \mathcal{S}^{-}_{mm_1'}\mathfrak{m}^+_{m'm_1} \right) - E^{(\beta)}_{\rm DC}.
\end{multline}
Here for $\beta = U$ we get from Eq.~\eqref{eq:W_U}
\begin{multline}
\mathcal{H}^{\rm HFA}_{\rm int}\left[W^{(U)}\right]
= 2U\sum_{m}
\left(
\mathcal{S}^{(0)}_{mm}\no_{mm} - 
\bm{\mathcal{S}}^z_{mm}\mathfrak{m}^z_{mm}  
\right) \\
- E^{(U)}_{\rm DC}
\end{multline}
and for $\beta = J_{\rm d}$ we get from Eq.~\eqref{eq:W_Jd}
\begin{multline}
\mathcal{H}^{\rm HFA}_{\rm int}\left[W^{(J_{\rm d})}\right]
= 2J_{\rm d}\sum_{m\ne m_1}
\left(
\mathcal{S}^{(0)}_{mm_1}\no_{mm_1} - 
\bm{\mathcal{S}}^z_{mm_1}\mathfrak{m}^z_{mm_1} 
\right) 
\\
- E^{(J_{\rm d})}_{\rm DC}.
\end{multline}
For the case $\beta = U', J_{\rm H}$ 
$W^{(\beta)}$ is spin independent, so we rewrite Eq.~\eqref{eq:H_int^HFA[W]}
%we use ??? for the first term in~Eq.~\eqref{eq:H_int^HFA[W]} and \eqref{eq:spin_combination_double} for the second one.
\begin{multline}\label{eq:H_int^HFA[W=U',J]}
\mathcal{H}^{\rm HFA}_{\rm int}\left[W^{(\beta)}\right]
= 2\sum_{mm'm_1m_1'}W^{(\beta)}_{\uparrow\downarrow}(mm';m_1m_1')
\\
\left(
2\mathcal{S}^{(0)}_{mm_1}\no_{m'm_1'} - 
\mathcal{S}^{(0)}_{mm_1'}\no_{m'm_1} - \bm{\mathcal{S}}_{mm_1'}\mob_{m'm_1} \right) \\
- E^{(\beta)}_{\rm DC},
\end{multline}
Here for $\beta = U'$ we get from Eq.~\eqref{eq:W_U'}
\begin{multline}
\mathcal{H}^{\rm HFA}_{\rm int}\left[W^{(U')}\right]
= 2U'\sum_{m\ne m'}
\left(
2\mathcal{S}^{(0)}_{mm}\no_{m'm'} - 
\right.
\\
\left.
- \mathcal{S}^{(0)}_{mm'}\no_{m'm} - \bm{\mathcal{S}}_{mm'}\mob_{m'm} \right) - E^{(U')}_{\rm DC},
\end{multline}
and for $\beta = J_{\rm H}$ we get from Eq.~\eqref{eq:W_J} 
\begin{multline}
\mathcal{H}^{\rm HFA}_{\rm int}\left[W^{(J_{\rm H})}\right]
= 2J_{\rm H}\sum_{m\ne m'}
\left(
2\mathcal{S}^{(0)}_{mm'}\no_{m'm} - 
\right.
\\
\left.
- \mathcal{S}^{(0)}_{mm}\no_{m'm'} - \bm{\mathcal{S}}_{mm}\mob_{m'm'} \right) - E^{(J_{\rm H})}_{\rm DC}.
\end{multline}

We write down the final  mean-field version of the Coulomb Hamiltonian as
\begin{equation}\label{eq:appendix:H_C_HF_final}
\mathcal{H}^{\rm HFA}_{\rm Coulomb} = 2\sum_{mm'} \left(\mathcal{F}^{(0)}_{mm'}\mathcal{S}^{(0)}_{mm'} - \bm{\mathcal{F}}_{mm'}\bm{\mathcal{S}}_{mm'}\right) - E_{\rm DC},
\end{equation}
where the contributions from four above-considered terms are collected together,
\begin{eqnarray}
\label{eq:appendix:f0_final}
\mathcal{F}^{(0)}_{mm'} &=& \delta_{mm'}\left[(2U'-J_{\rm H})K + \delta U \no_{mm}\right] 
\\
&+& (2J_{\rm H} - U')\no_{m'm} + J_{\rm d}\no_{mm'} , \\
\label{eq:appendix:f_vec_final}
\bm{\mathcal{F}}_{mm'} &=& \delta_{mm'}\left[J_{\rm H}\mathbf{M} + \delta U\mob_{mm}\right] 
\\
&+& U'\mob_{m'm} + J_{\rm d}\mob_{mm'}, \\
\label{eq:appendix:E_DC_final}
E_{\rm DC} &=& \sum_{mm'} \left(\mathcal{F}^{(0)}_{mm'}\no_{mm'} - \bm{\mathcal{F}}_{mm'}\mob_{mm'}\right),
\end{eqnarray}
where $\delta U = U- U' - J_{\rm H} - J_{\rm d}$, where 
%$K$ and $\mathbf{M}$ are defined by Eqs.~(\ref{eq:K_def}) and (\ref{eq:M_def}). 
\begin{eqnarray}
\label{eq:K_def}
K &=& \sum_m\no_{mm},\\
\label{eq:M_def}
\mathbf{M} &=& \sum_m\mob_{mm}.
\end{eqnarray}
Within the Kanamori approximation $\delta U = 0$ and we employ this in the main text.

\subsection{Derivation of HFA Hamiltonian}
Below we present  details of the HFA approximation for the Hamiltonian~\eqref{eq:hamilt}, using the results of Sec. 1 of this appendix, restoring the site index $i$.

We apply the transformation to the Bloch wave functions in~Eq.~(\ref{eq:C_local_def}),
\begin{equation}\label{eq:from_Wannier_to_Blokh}
	c_{im\sigma} = N^{-1/2}\sum_{\mathbf{k}}\exp(-\I\mathbf{k}\mathbf{R}_i)c_{\mathbf{k}m\sigma},
\end{equation}
and the Fourier transform of $C^i_{m\sigma;m'\sigma'}$ turns out to be
\begin{equation}\label{eq:C_Fourier}
C_{m\sigma;m'\sigma'}(\mathbf{q}) = \frac1{N}\sum_{\mathbf{k}}\langle c^\dag_{\mathbf{k}m\sigma}c^{}_{\mathbf{k}+\mathbf{q},m'\sigma'}\rangle.
\end{equation}

The result of application of HFA to the Hamiltonian~(\ref{eq:main_H_Coulomb}) is 
\begin{multline}
\mathcal{H}^{\rm MF}_{\rm Coulomb} = \sum_{\mathbf{k}\mathbf{k}'}\sum_{\sigma\sigma'}\sum_{mm'}\left(\mathcal{F}^{(0)}_{mm'}(\mathbf{k}-\mathbf{k}')\delta_{\sigma\sigma'} 
\right.
\\
- 
\left.
\bm{\mathcal{F}}_{mm'}(\mathbf{k}-\mathbf{k}')\bm{\sigma}_{\sigma\sigma'}
\right)c^\dag_{\mathbf{k}m\sigma}c^{}_{\mathbf{k}'m'\sigma'},
\end{multline}
where mean fields $\mathcal{F}^{(0)}_{mm'}(\mathbf{k}-\mathbf{k}')$ and $\bm{\mathcal{F}}_{mm'}(\mathbf{k}-\mathbf{k}')$ are Fourier transforms of mean fields given by Eqs.~\eqref{eq:Fn^i_def} and \eqref{eq:Fm^i_def}. 
%Here we define averages 
%Analogously to Eqs.~(\ref{eq:appendix:n_mm_def}), (\ref{eq:appendix:m_mm_def}), , we define the matrices 
%$\no_{mm'}(\mathbf{q}) = c^{(0)}_{mm'}(\mathbf{q}), \mob_{mm'}(\mathbf{q}) = \mathbf{c}_{mm'}(\mathbf{q})$, where .
%Here the definitions 
%\begin{eqnarray}
%\label{eq:K_def}
%K(\mathbf{q}) &=& \sum_m\no_{mm}(\mathbf{q}),\\
%\label{eq:M_def}
%\mathbf{M}(\mathbf{q}) &=& \sum_m\mob_{mm}(\mathbf{q})
%\end{eqnarray}
%are used.

The treatment of the Hamiltonian \eqref{eq:H_MF_total} is presented below. 
We introduce the magnetic Brillouin zone $(|k_x| + |k_y|< \pi)$, so that for any $\mathbf{k}_1$ from the Brillouin zone we have a presentation $\mathbf{k}_1 = \mathbf{k} + \alpha\mathbf{Q}$, where $\alpha = 0, 1$. We rewrite Eq.~\eqref{eq:H_MF_total} through the summation over the magnetic Brillouin zone (denoted by a prime)
\begin{equation}
\mathcal{H}_{\rm MF} = \sum_{\mathbf{k}\alpha\alpha';mm'}'\sum_{\sigma\sigma'}H^{\rm MF}_{m\sigma\alpha;m'\sigma'\alpha'}(\mathbf{k})c^\dag_{\mathbf{k} + \alpha\mathbf{Q},m\sigma}c^{}_{\mathbf{k} + \alpha'\mathbf{Q},m'\sigma'},
\end{equation}
%Introducing the $\mathbf{k}$ summation over magnetic Brillouin zone
where the $\mathbf{k}$-dependent $3\times2\times2$ matrix 
\begin{multline}\label{eq:H_matrix}
H^{\rm MF}_{m\alpha\sigma;m'\alpha'\sigma'}(\mathbf{k}) = \left[
\left(\varepsilon_{mm'}(\mathbf{k} + \alpha\mathbf{Q}) + \Delta_{\rm CF}\delta_{m,xy}\right)\delta_{mm'}\delta_{\sigma\sigma'} 
\right.
\\
\left.
+ \mathcal{F}^{(0)\rm u}_{mm'}\delta_{\sigma\sigma'} - {\bm{\mathcal{F}}^{\rm u}_{mm'}}\bm{\sigma}_{\sigma\sigma'} - (\lambda/2)\mathbf{l}_{mm'}\bm{\sigma}_{\sigma\sigma'}
\right]\delta_{\alpha\alpha'} 
\\
+ \left[\mathcal{F}^{(0)\rm s}_{mm'}\delta_{\sigma\sigma'} - \bm{\mathcal{F}}^{\rm s}_{mm'}\bm{\sigma}_{\sigma\sigma'}\right]\delta_{\alpha\bar\alpha'},
\end{multline}
is introduced. Here $\varepsilon_{mm'}(\mathbf{k})$ sets up the band spectrum, the explicit expression for which is given in the end of Sec.~\ref{sec:model_method}.

The Hamiltonian matrix~\eqref{eq:H_matrix} is diagonalized by the transformation (cf.~\cite{Igoshev:2010}) 
\begin{equation}
	c_{\mathbf{k}+\alpha\mathbf{Q},m\sigma} = \sum_{\nu}a_{m\alpha\sigma;\nu}(\mathbf{k})d_{\mathbf{k}\nu}.
\end{equation}
In terms of new Fermi operators $d_{\mathbf{k}\nu}$, $d_{\mathbf{k}\nu}^\dag$ the Hamiltonian has the form
\begin{equation}
\mathcal{H}_{\rm MF} = \sum_{\mathbf{k}\nu}'E_{\nu}(\mathbf{k})d^\dag_{\mathbf{k}\nu}d^{}_{\mathbf{k}\nu}.
\end{equation}
Then the correlator (\ref{eq:C_Fourier}) can be expressed through the spectrum  $E_{\nu}(\mathbf{k})$ and coefficients  $a_{m\alpha\sigma;\nu}(\mathbf{k})$:
\begin{multline}\label{eq:order_k_form_alpha}
\langle c^\dag_{\mathbf{k}+\alpha\mathbf{Q},m\sigma}c^{}_{\mathbf{k}'+\alpha'\mathbf{Q},m'\sigma'}\rangle \\
= \delta_{\mathbf{k}\mathbf{k}'}\sum_{\nu}a^*_{m\alpha\sigma;\nu}(\mathbf{k})a_{m'\alpha'\sigma';\nu}(\mathbf{k})f(E_\nu(\mathbf{k})).
\end{multline}
Applying Eq.~\eqref{eq:appendix:E_DC_final} we obtain 
\begin{multline}\label{eq:E_DC_final}
E_{\rm DC}/N = \sum_{\mathbf{q}=0,\mathbf{Q}}\left[(2U' - J_{\rm H})K^2(\mathbf{q}) - J_{\rm H}\mathbf{M}^2(\mathbf{q}) 
\right. \\ \left.
+ \delta U\sum_{m}\left(\no^2_{mm}(\mathbf{q}) - \mob^2_{mm}(\mathbf{q})\right) 
\right. \\ \left.
+
J_{\rm d}\sum_{mm'}\left(\no^2_{mm'}(\mathbf{q}) - \mob^2_{mm'}(\mathbf{q})\right) 
\right. \\ \left.
- (U' - 2J_{\rm H} )\sum_{mm'}\no_{mm'}(\mathbf{q})\no_{m'm}(\mathbf{q}) 
\right. \\ \left.
- U'\sum_{mm'}\mob_{mm'}(\mathbf{q})\cdot\mob_{m'm}(\mathbf{q})\right].
\end{multline}

\end{document}